\documentclass[aps,prb,preprint,showpacs,preprintnumbers,amsmath,amssymb]{revtex4-1}

\usepackage{graphicx}
\usepackage{epstopdf}
\usepackage{comment}
\usepackage{tabularx}
\usepackage{amsmath}

\begin{document}

\preprint{}

\title{"Head-to-head" and "tail-to-tail" $180^\circ$ domain walls in an isolated ferroelectric}

\author{M. Y. Gureev}
\email{maxim.gureev@epfl.ch}
\author{A. K. Tagantsev}
\author{N. Setter}
\affiliation{Ceramics Laboratory, Swiss Federal Institute of Technology (EPFL), CH-1015 Lausanne, Switzerland}
\date{\today}

\begin{abstract}
ÓHead-to-headÓ and Ótail-to-tailÓ $180^\circ$ domain-walls in a finite isolated ferroelectric sample are theoretically studied using Landau theory.
The full set of equations, suitable for numerical calculations is developed.
The explicit expressions for the polarization profile across the walls are derived for several limiting cases and wall-widths are estimated.
It is shown analytically that different regimes of screening and different dependences for width of charged domain walls on the temperature and parameters of the system are possible, depending on spontaneous polarization and concentration of carriers in the material.
It is shown that the half-width of charged domain walls in typical perovskites is about the nonlinear Thomas-Fermi screening-length and about one order of magnitude larger than the half-width of neutral domain-walls.
The formation energies of Óhead-to-headÓ walls under different regimes of screening are obtained, neglecting the poling ability of the surface.
In the nonlinear regimes of screening, this energy is equal to the energy necessary for the creation of electron-hole pairs in the amount sufficient to screen the spontaneous polarization, which is proportional to the band gap of the ferroelectric.
It is shown that either Òhead-to-headÓ or Òtail-to-tailÓ configuration can be energetically favorable in comparison with the monodomain state of the ferroelectric if the poling ability of the surface is large enough.
If this is not the case, the existence of charged domain walls in bulk ferroelectrics is merely a result of the domain-growth kinetics.
Formation energies of the other possible states: multidomain state with antiparallel domains separated by neutral walls and the state with the zero polarization were compared with the formation energy of the charged domain wall.
It was shown that, at large enough sample thicknesses, a charged domain wall can be energetically favorable in comparison with the states mentioned above.
This size effect could explain why charged domain walls were observed experimentally in bulk lead titanate but not in barium titanate.
The results obtained for the case of an isolated ferroelectric sample were compared with the results for an electroded sample.
It was shown that charged domain wall in electroded sample can be either metastable or stable, depends on the work function difference between electrodes and ferroelectric and the poling ability of the electrode/ferroelectric interface.
\end{abstract}

\maketitle

\section{Introduction}\label{Introduction}

Ferroelectric materials are widely used today in memories, piezoelectric transducers, pyroelectric detectors and thin-film capacitors\cite{Setter,Scott}. The structural and functional properties of domain walls can substantially influence poling and polarization switching - two most important processes in ferroelectrics and their applications.

\begin{figure}[h]
\begin{center}
\includegraphics[width=0.7\textwidth]{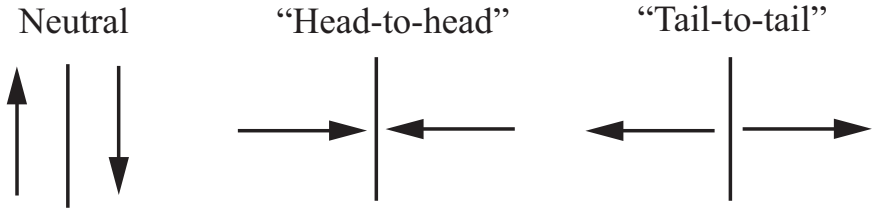}
\caption{Neutral and two types of charged domain walls, with the polarization normal to the wall surface. Orientation of spontaneous polarization with respect to the wall is shown with arrows.}
\label{configurations}
\end{center}
\end{figure}

Three qualitatively different configurations are possible for 180-degrees domains (Fig. \ref{configurations}). In the first case, the domain wall is parallel to the direction of the spontaneous polarization inside the adjacent domains. In this configuration there is no bound charge on the wall. This case is well studied and described in details (see e.g. the book by Strukov and Levanyuk\cite{Levanyuk}).
In the other two configurations, the domain wall is not parallel to the direction of the spontaneous polarization.
The polarization can be directed either towards the domain wall ("head-to-head") or from the domain wall ("tail-to-tail") so that positive or negative bound charges are present on the domain wall, respectively.

Although neutral domain walls are much more common, charged domain walls were observed in various ferroelectrics,
for example in PbTiO$_3$ crystals\cite{Fesenko73,Surowiak,Fesenko85},
Pb$[$Zr$_x$Ti$_{1-x}]$O$_3$ (PZT) ceramics\cite{Randall}, BiFeO$_3$\cite{Ramesh}, and PZT thin films\cite{Jia}.

The properties of charged domain walls can be quite different from the properties of neutral domain walls.
For example Mokry, Tagantsev, and Fousek\cite{Mokry} showed that the compensation of the polarization charge by free carriers reduces the pressure excerted on the wall when an external electric field is applied.
This leads to a reduction in the mobility of the wall and an increase in switching voltage, even when the mobility of the compensating free charge is high. This effect was experimentally observed by Balke et. al\cite{Kalinin} in BiFeO$_3$ thin film.

In a perfect insulator, unscreened bound charge on the charged domain wall typically creates a very large electric field,
 which effectively shifts down the Curie temperature (for the wall perpendicular to the direction of polarization in barium titanate this shift is about 24000 K.)
Therefore, except for a very small spontaneous polarization and a very small angle between the vector of spontaneous polarization and the domain wall, for charged walls to exist, their bound charge should be almost completely screened by free charges.
These free charges should be taken into consideration in the theory for charged domain walls.
It is this situation that will be addressed in the paper.
If it is not mentioned specifically, we will consider "normal" ferroelectrics, with high permittivity $\varepsilon_\mathrm{f}\gg\varepsilon_\mathrm{b}$, where $\varepsilon_\mathrm{b}$ is the background permittivity.
The model considered is suitable only for the case of the big angle between the domain wall and the vector of spontaneous polarization.

"Head-to-head" and "tail-to-tail" domain walls in a sample with metallic electrodes were considered in a series of works by Ivanchik, Guro, Chenskii et al.\cite{Chenskii,Guro68,Guro70,Vul}
It was argued that such domain walls can be stable depending on the work function difference between the metal electrode and the ferroelectric.
Below we apply the Landau theory to address the problem of charged domain walls in an isolated ferroelectric.
Some models of "head-to-head" and "tail-to-tail" domain walls in an isolated sample were considered in the past,
where the wall was stabilized by an inhomogeneous distribution of a dopant.\cite{Yudin,Vanderbilt}
We will consider the case of a homogeneous material.
The results obtained below for the internal structure and characteristic scales of domain walls can also be directly applied to the case of an electroded sample as well.
We will revisit the problem of domain wall width because several different scales were obtained in the past by different authors.
For example, according to the classical book on ferroelectric-semiconductors by Fridkin\cite{Fridkin}, domain wall formation in head-to-head configuration is controlled by Debye screening with free carriers and the wall thickness should be about the Debye screening length. Another scale was found by Krapivin and Chenskii.\cite{Chenskii}
We also compare the results obtained for the isolated ferroelectric with the results obtained in the past for the electroded samples. \cite{Chenskii,Guro70,Vul}

The paper develops as follows:
First, we formulate the problem in a way suitable for numerical calculations (Sect. \ref{General}).
In Sect. \ref{Scales} we discuss characteristic scales of the polarization variation and limits of applicability of the continuos theory.
In Sect. \ref{Analytical} the problem is formulated in a way, suitable for an analytical treatment.
We obtain analytical solutions and evaluate domain wall widths corresponding to the different regimes of polarization screening by free carriers (Sect. \ref{Classic} and \ref{Degenerate}).
The correspondence between the regimes of screening and parameters of the ferroelectric is discussed in Sect. \ref{Comparison}.

In Sect. \ref{Energies} we evaluate the energies of isolated samples with "head-to-head" and "tail-to-tail" domain walls.
Different configurations that can be energetically favorable depending on the thickness of the ferroelectric, and size-effects are considered in Sect. \ref{Size}.
An analysis is presented in Sect. \ref{electrodes}, where the situation in isolated ferroelectric sample is compared with that in an electroded sample.
Numerical estimates and comparison of the prediction of the theory with experimental data are presented in Sect. \ref{Experiment}.

\section{General formulation of the problem}\label{General}

We consider "head-to-head" and "tail-to-tail" domain walls perpendicular to the direction of polarization (as shown in Fig. \ref{configurations}) in an isolated sample.
If the ferroelectric material is an ideal insulator, a huge depolarizing field will appear, which will suppress ferroelectricity in the sample.
As a model, we will consider an infinite plate with thickness $L$.
We consider a single component system, where all the variables depend only on one $x$ coordinate, where the $x$ axis is perpendicular to the plate surface.

First, we obtain the system of equations for the isolated ferroelectric element using the Landau theory, taking into account the screening charges.
For simplicity we consider ferroelectrics with a second order phase transition.
However, most of the results obtained below are valid for ferroelectrics with a first order phase transition as well.

The equation of state for ferroelectrics with a second order phase transition has the form (see e.g. \cite{Tagantsev_book}):
\begin{equation}\label{state}
E=\alpha P+\beta P^3-\kappa\left(\frac{\partial^2P}{\partial x^2}\right)
\end{equation}
where $\alpha<0$ in the ferroelectric state, $\beta>0$, and $\kappa$ is the coefficient of the gradient term in the free energy, $E$ is the electric field, and $P$ is the ferroelectric part of the polarization.

The electrical field and potential $\varphi$ are connected by the relation:
\begin{equation}\label{E}
E=-\frac{\partial\varphi}{\partial x}.
\end{equation}

The Poisson equation for one-dimensional case has the form:
\begin{equation}\label{puasson}
 \frac{\partial D}{\partial x} = 4\pi \rho
\end{equation}
where $\rho$ is the free charge density, $D$ is the electrical displacement, defined as:
\begin{equation}\label{displacement}
D=\varepsilon_\mathrm{b}E+4\pi P,
\end{equation}
 $\varepsilon_\mathrm{b}$ is the background permittivity.

In the model of electron gas with parabolic spectrum the free charge density $\rho$  depends on the potential $\varphi$ as follows\cite{Sze,Liu,Xiao}:
\begin{equation}\label{charge_density}
\rho=-qN_\mathrm{C}F_\mathrm{1/2}\left(\frac{E_\mathrm{F}-E_\mathrm{C}+q\varphi}{kT}\right)
+qN_\mathrm{V}F_\mathrm{1/2}\left(\frac{E_\mathrm{V}-E_\mathrm{F}-q\varphi}{kT}\right)
+qz_\mathrm{d}N_\mathrm{d}t_\mathrm{d}(\varphi)-qz_\mathrm{a}N_\mathrm{a}t_\mathrm{a}(\varphi);
\end{equation}
$N_\mathrm{C}$ and $N_\mathrm{V}$ are the effective density of states in conductive and valence band respectively,
$E_\mathrm{V}$ is the top of the valence band and $E_\mathrm{C}$ is the bottom of the conduction band,
$E_\mathrm{F}$ is the Fermi level, $q$ is the absolute value of the electron charge,  $T$ is the absolute temperature and $k$ is the Boltzmann constant,
$F_\mathrm{1/2}$ is the Dirac-Fermi integral, defined as:
\begin{equation}\label{fermi_integral}
F_\mathrm{1/2}(\sigma)=\frac{2}{\sqrt{\pi}}\int\limits_0^\infty\frac{\xi^{1/2}d\xi}{1+\mathrm{exp}(\xi-\sigma)},
\end{equation}
$z_\mathrm{a}$ and $z_\mathrm{d}$ are the acceptor and donor valencies, $t_\mathrm{d}$ and $t_\mathrm{a}$ are the fraction of ionized donors and acceptors respectively given by:

\begin{equation}\label{ionized_donors}
t_\mathrm{d}(\varphi)=1-\frac{1}{1+\frac{1}{g_\mathrm{d}}\exp\left(\frac{E_\mathrm{d}-E_\mathrm{F}-q\varphi}{kT}\right)},
\end{equation}

\begin{equation}\label{ionized_acceptors}
t_\mathrm{a}(\varphi)=\frac{1}{1+g_\mathrm{a}\exp\left(\frac{E_\mathrm{a}-E_\mathrm{F}-q\varphi}{kT}\right)},
\end{equation}
where $E_\mathrm{d}$ and $E_\mathrm{a}$ are the donor and acceptor level respectively, $g_\mathrm{d}$ is the ground state degeneracy of the donor impurity level and $g_\mathrm{a}$ is the degeneracy of acceptor level.
As we will show later, the carriers concentration in a charged domain wall is much smaller than the half-band filling one, and so that the use of the parabolic approximation is justified.

The isolated ferroelectric element is electrically neutral. There is no electric field outside the sample, thus the boundary conditions can be formulated as follows:
\begin{equation}\label{boundary_D}
D|_{x=-L/2}=D|_{x=L/2}=0
\end{equation}
The other pair of boundary conditions given as (see e.g. \cite{Glinchuk}):
\begin{equation}\label{boundary_surface}
\left(\left.\kappa\frac{\partial P}{\partial x}+\zeta\mp\eta P\right)\right|_{x=\pm L/2}=0
\end{equation}
where $\zeta$ and $\eta$ are coefficients in the surface free energy expansion with respect to $P$.

Equations \eqref{state}-\eqref{charge_density} constitute the full set of equations for the problem.
This set of equations with boundary conditions Eqs. \eqref{boundary_D}, \eqref{boundary_surface} and additional conditions
\begin{equation}\label{head-to-head-cond}
\begin{cases} P>0, & \mbox{if } -L/2<x<0 \\ P<0, & \mbox{if } 0<x<L/2 \end{cases}
\end{equation}
\begin{equation}\label{head-to-head-cond}
\begin{cases} P<0, & \mbox{if } -L/2<x<0 \\ P>0, & \mbox{if } 0<x<L/2 \end{cases}
\end{equation}
 can be solved numerically to obtain the polarization distribution in the sample with "head-to-head" and "tail-to-tail" domain wall respectively. Due to the symmetry of the problem boundary conditions can be also redefined as $P(0)=0$, $D(0)=0$ and Eqs. \eqref{boundary_D}, \eqref{boundary_surface} at $x=L/2$.

We define the domain wall formation energy as:
\begin{equation}\label{DWenergy}
W=\int_{-L/2}^{L/2}[ \Phi(x)-\Phi_0(x)]dx+\Phi_\mathrm{surf}
\end{equation}
where $\Phi(x)$ is the free energy density of a system with the domain wall and
$\Phi_0(x)=\frac{\alpha}{2}P_0^2+\frac{\beta}{4}P_0^4+\Phi_\mathrm{eg}(n_0)=\frac{\alpha}{4}P_0^2+\Phi_\mathrm{eg}(n_0)$, $P_0=\sqrt{-\alpha/\beta}$, $\Phi_\mathrm{eg}(n_0)$
is the electron gas free energy at equilibrium concentration of carriers in the homogeneous material.
Hereafter we will use a shorthand "homogeneous concentration" for this way defined carrier concentration.
The integration is done over the thickness of the sample. The $\Phi_\mathrm{surf}$ is the surface energy per unit area given as follows \cite{Minyukov,Tagantsev}:

\begin{equation}\label{Surface_energy}
\Phi_\mathrm{surf}=\zeta P|_{x=-L/2}-\zeta P|_{x=L/2}+\frac{\eta}{2} P^2|_{x=-L/2}+\frac{\eta}{2} P^2|_{x=L/2}
\end{equation}

The free energy density of the system consists of the free energy of the lattice, the electrostatic energy, and the free kinetic energy density of the electron gas $\Phi_\mathrm{eg}$:
\begin{equation}\label{fe}
 \Phi = \frac{\alpha}{2}P^2 + \frac{\beta}{4}P^4+\frac{\kappa}{2}\left(\frac{\partial P}{\partial x}\right)^2+\frac{\varepsilon_bE^2}{8\pi}+\Phi_\mathrm{eg}.
\end{equation}

The free energy density of the electron gas can be defined as:
\begin{equation}\label{freeenergy}
\Phi_\mathrm{eg}=E_\mathrm{eg}-TS_\mathrm{eg}
\end{equation}
where $E_\mathrm{eg}$ is the kinetic energy density of the electron gas and $S_\mathrm{eg}$ is its entropy density.
The kinetic energy can be found by integration over the valence and conduction bands and impurities:
\begin{equation}\label{energy}
E_\mathrm{eg}=\int\limits^{E_\mathrm{V}}_{-\infty}E N_\mathrm{V}(E) f(E)dE+\int\limits_{E_\mathrm{C}}^{\infty}E N_\mathrm{C}(E) f(E)dE
+z_\mathrm{a}N_\mathrm{a}t_\mathrm{a}(\varphi)E_\mathrm{a}+z_\mathrm{d}N_\mathrm{d}(1-t_\mathrm{d}(\varphi))E_\mathrm{d}
\end{equation}
where $N_\mathrm{V}(E)$ and $N_\mathrm{C}(E)$ are the density of states in valence and conduction bands respectively, $f(E)$ is the Fermi function.
For the case of parabolic spectrum this relation can be transformed to the form:
\begin{equation}\label{energy_bandgap}
E_\mathrm{eg}=nE_\mathrm{V}+n_\mathrm{C}E_\mathrm{g}
+N_\mathrm{C}F_\mathrm{3/2}\left(\frac{E_\mathrm{F}-E_\mathrm{C}+q\varphi}{kT}\right)-N_\mathrm{V}F_\mathrm{3/2}\left(\frac{E_\mathrm{V}-E_\mathrm{F}-q\varphi}{kT}\right)
+z_\mathrm{a}N_\mathrm{a}t_\mathrm{a}(\varphi)E_\mathrm{a}-z_\mathrm{d}N_\mathrm{d}t_\mathrm{d}(\varphi)E_\mathrm{d}
\end{equation}
where $E_\mathrm{g}$ is the bandgap and $F_\mathrm{3/2}$ defined as:

\begin{equation}\label{F_3_2}
F_\mathrm{3/2}(\sigma)=\frac{2}{\sqrt{\pi}}\int\limits_0^\infty\frac{\xi^{3/2}d\xi}{1+\mathrm{exp}(\xi-\sigma)}.
\end{equation}

The entropy density for Fermi gas\cite{Landau5} can be found as follows:
\begin{equation}\label{entropy}
S_\mathrm{eg} =-k\sum_{i}\left[f_i\ln f_i+\left(1-f_i\right)\ln \left(1-f_i\right)\right].
\end{equation}
where summation is done over all possible electronic states, $f_i$ is the occupation probability for the $i$-th state.

%%%%%%%%%%%%%%%%%%%%%%%%%%%%

\section{Limits of applicability of the continuos theory and analytical solutions in a large crystal}\label{Analytical_Solutions}

\subsection{Characteristic scales and limits of applicability of the continuos theory}\label{Scales}

In the previous section we derived the full set of equations, which can be solved numerically to obtain the exact solution of the problem.
There exist the number of situations where approximate analytical solutions to this set can be obtained.
Below we will introduce some approximations to within a small parameter  $\varepsilon_\mathrm{b}/\varepsilon_\mathrm{f}$, where $\varepsilon_\mathrm{f}=2\pi/|\alpha|$ is the contribution of the ferroelectric subsystem to the permittivity of the material, which enable us to get these solutions.

Using Eq. \eqref{displacement}, the Poisson equation can be presented as:
\begin{equation}\label{poissin_E}
\varepsilon_\mathrm{b}\frac{\partial E}{\partial x}+4\pi \frac{\partial P}{\partial x}=4\pi\rho.
\end{equation}

After differentiation of Eq. \eqref{poissin_E} and using Eq. \eqref{E} one can get:
\begin{equation}\label{poissin_Edif}
\frac{\partial^2 E}{\partial x^2}+\frac{4\pi}{\varepsilon_\mathrm{b}}\frac{\partial \rho}{\partial \varphi}E=-\frac{4\pi}{\varepsilon_\mathrm{b}}\frac{\partial^2P}{\partial x^2}.
\end{equation}
Equation \eqref{poissin_Edif} can be rewritten as:
\begin{equation}\label{poissin_Edif_l0}
\frac{\partial^2 E}{\partial x^2}+\frac{1}{l_0^2}E=-\frac{4\pi}{\varepsilon_\mathrm{b}}\frac{\partial^2P}{\partial x^2},
\end{equation}
where
\begin{equation}\label{l_0}
l_0=\sqrt{-\frac{\varepsilon_\mathrm{b}}{4\pi(\partial \rho/\partial \varphi)}}.
\end{equation}

This equation contains two self-consistent characteristic scales for the spatial variation of the polarization. The big one corresponds to the case, where the first term in Eq. \eqref{poissin_Edif_l0} can be neglected, and this equation can be presented as:
\begin{equation}\label{poissin_big}
\frac{\partial \rho}{\partial \varphi}E=\frac{\partial^2P}{\partial x^2}.
\end{equation}
Substituting equation of state Eq. \eqref{state} into Eq. \eqref{poissin_big} one can get:
\begin{equation}\label{big_scale_general}
\frac{\partial \rho}{\partial \varphi}\left(\alpha P+ \beta P^3-\kappa\left(\frac{\partial ^2P}{\partial x^2}\right)\right)=\frac{\partial^2P}{\partial x^2}.
\end{equation}
The characteristic scale $\delta$ can be estimated from Eq. \eqref{big_scale_general} as:

\begin{equation}\label{big_scale_rc}
\delta^2\approx\frac{-1}{2|\alpha|(\partial\rho/\partial\varphi)}+r_\mathrm{c}^2,
\end{equation}
where $r_c$ is the correlation radius, defined as:
 \begin{equation}\label{r_c}
r_\mathrm{c}=\sqrt{\frac{\kappa}{2|\alpha|}}.
\end{equation}
Hereafter in the estimates, $\partial\rho/\partial\varphi$ means typical value of $\partial\rho/\partial\varphi$ inside the region with pronounced variation of polarization (e. g. domain wall).

As we will see later, for any realistic situation $\delta\gg r_\mathrm{c}$, thus from Eq. \eqref{big_scale_rc} one can get:
\begin{equation}\label{big_scale}
\delta\approx\sqrt{\frac{-1}{2|\alpha|(\partial\rho/\partial\varphi)}}.
\end{equation}
To check the self-consistency of this scale we should show that when the spatial variations of the polarization is controlled by this scale the first term in Eq. \eqref{poissin_Edif_l0} is small in comparison with the second one, i.e. $\delta\gg l_0$.
From Eq. \eqref{l_0} and Eq. \eqref{big_scale} we can estimate the ratio of these terms to be about:
\begin{equation}\label{big_scale_to_l0}
\frac{\delta^2}{l_0^2}\approx\frac{\varepsilon_\mathrm{f}}{\varepsilon_\mathrm{b}}\gg 1.
\end{equation}
Thus our assumption $\delta\gg l_0$ is self-consistent.
This result is also supported by more rigorous calculations as demonstrated  in App. \ref{appendix}, using the exact solutions to  Eq. \eqref{big_scale_general}.

\begin{figure}[h]
\begin{center}
\includegraphics[width=0.7\textwidth]{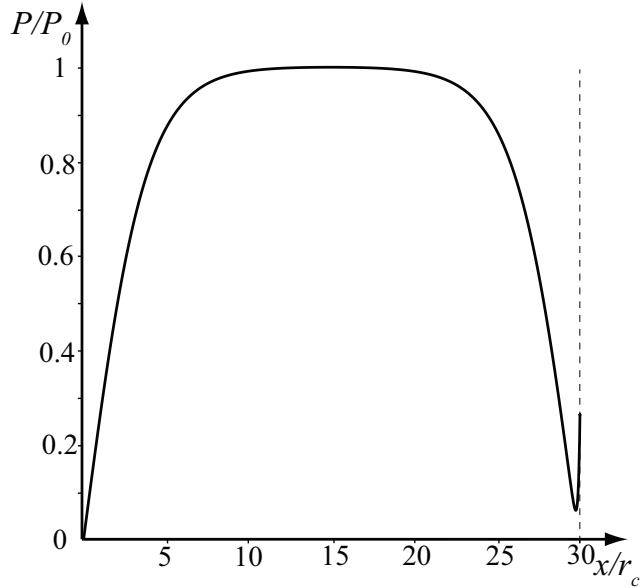}
\caption{Numerical solution with boundary conditions Eqs. \eqref{boundary_surface}, \eqref{boundary_D}, and $P(0)=0$, $D(0)=0$. The small parameter $\varepsilon_\mathrm{b}/\varepsilon_\mathrm{f}=0.01$, $\delta/r_\mathrm{c}=4$. The dimensionless surface energy coefficients $\eta r_\mathrm{c}/\kappa=5000$, $\zeta r_\mathrm{c}/(\kappa P_0)=5000$.}
\label{num_bound}
\end{center}
\end{figure}

As the result of the simplification we have passed from a set of the equations equivalent to a fourth order differential equation to one second order equation, Eq.\eqref{big_scale_general}.
At the same time we still have 4 boundary conditions, Eqs. \eqref{boundary_D} and  \eqref{boundary_surface}. 
Thus we have  excess of the boundary conditions.
A problem of such kind is a  known in theoretical physics, for example, in hydrodynamics of liquid flowing in a tube \cite{Landau6}.
In that case, the problem  was solved by introducing the boundary layer, the description of which goes beyond the simplified theory.
We adapt a similar approach by assuming  that there exist a thin layer near the surface, where our simplified scheme does not work and where the charge density is small in comparison with the $\partial P/\partial x$.
In this case Eq. \eqref{poissin_E} together with Eq. \eqref{boundary_D} leads to the following equation:
\begin{equation}\label{Dequal0}
\varepsilon_\mathrm{b}E+4\pi P=0.
\end{equation}
Using equation of state Eq. \eqref{state} one can rewrite it as:
\begin{equation}\label{poissin_small1}
\kappa\frac{\partial^2 P}{\partial x^2}+\alpha P+\beta P^3+\frac{4\pi}{\varepsilon_\mathrm{b}} P=0.
\end{equation}
Neglecting terms about $\varepsilon_\mathrm{b}/\varepsilon_\mathrm{f}$, the characteristic thickness of the boundary layer can be found from Eq. \eqref{poissin_small1} as follows:
\begin{equation}\label{boundary_l}
l=\sqrt{\frac{\varepsilon_\mathrm{b}\kappa}{4\pi}}=\sqrt{ \frac{\varepsilon_\mathrm{b}}{\varepsilon_\mathrm{f}} }r_\mathrm{c}.
\end{equation}
We can shift boundary condition $D=0$ on a distance $l$ from the surface, and the polarization changes on this scale to satisfy the second boundary condition.
The condition $D=0$ means that we neglect the charge in the surface layer, that is exactly the assumption we made to obtain scale $l$, thus the solution with scale $l$ is self-consistent.
Such approach can be supported by numerical calculations.
The numerical solution obtained for the case of linear screening with classical gas with exact boundary conditions is shown on Fig. \ref{num_bound}.
Here it is seen a deep minimum, which is corresponds to the border of the surface boundary layer.

For typical perovskites Eq. \eqref{boundary_l} implies $l\approx\sqrt{\kappa}\approx a$, where $a$ is the lattice constant.
Thus the continuous theory is not applicable inside this surface layer.
This means that  the result of $l$ obtained above can be considered as an indication that $l$ is atomically small.
The numerical solutions obtained for the full set of equations with full boundary conditions will also contain the scale smaller than the lattice constant, thus this solution near the surface will be outside of the applicability of the continuos theory.
Only the solutions with big scale and shifted boundary condition can be used.
The consideration of the properties of the surface layer requires an ab initio approach.
For some materials, like week ferroelectrics \cite{Tagantsev87} or ferroelectrics of order-disorder type far from the phase transition $\varepsilon_\mathrm{b}/\varepsilon_\mathrm{f}$ can readily be of the order of one or larger.
In this case the inequality \eqref{big_scale_to_l0} is not valid anymore, and the approximation of the boundary layer becomes not self-consistent.
However, at the same time, the small scale $l$ becomes macroscopic, of the order of the correlation radius,
thus the solution can be found by solving full system with exact boundary conditions,
without leaving the range of applicability of the continuous theory.

Below we will consider normal ferroelectrics (with $\varepsilon_\mathrm{b}/\varepsilon_\mathrm{f}\ll 1$ ).
We will not consider the structure of the surface layer and will use macroscopic boundary condition $D=0$.

\subsection{Problem formulation for analytical solution}\label{Analytical}

Now that the boundary conditions are defined, we can calculate the polarization distribution in the sample.
Equation \eqref{big_scale_general} can be rewritten as:
\begin{equation}\label{big_scale_combined}
\alpha P+\beta P^3=\left(\kappa-(\partial \rho/\partial \varphi)^{-1}\right)\frac{\partial^2P}{\partial x^2}.
\end{equation}
This is an equation with respect to $P$, and it is convenient to reformulate the boundary conditions with respect to $P$ instead of $D$.
From Eq. \eqref{big_scale_rc} it is immediately follows $\delta\geq r_\mathrm{c}$.
 With Eqs. \eqref{state}, \eqref{displacement} one can get:
\begin{equation}\label{DP}
D=\varepsilon_\mathrm{b}\left(\alpha P+\beta P^3+\kappa\frac{\partial^2 P}{\partial x^2}\right)+4\pi P
\end{equation}
In out problem $|P|\leq |P_0|$ that leads to $|\beta P^3|\leq |\alpha P|$, the gradient term $|\kappa(\partial^2 P/\partial x^2)|\leq|\kappa(P/r_\mathrm{c}^2)|\approx|\alpha P|$, thus in the lowest approximation with respect to $\varepsilon_\mathrm{b}/\varepsilon_\mathrm{f}\ll 1$, Eq. \eqref{DP} can be rewritten as:
\begin{equation}\label{D4piP}
D=4\pi P.
\end{equation}
So that boundary condition Eq. \eqref{big_scale_solution_r_c} can be presented as:
\begin{equation}\label{Boundary_P}
P|_{x=\pm L/2}=0.
\end{equation}
The Poisson equation, Eq. \eqref{puasson}, in this case can be rewritten as:
\begin{equation}\label{Poisson_P}
\frac{\partial P}{\partial x}=\rho.
\end{equation}

\begin{figure}[h]
\begin{center}
\includegraphics[width=0.7\textwidth]{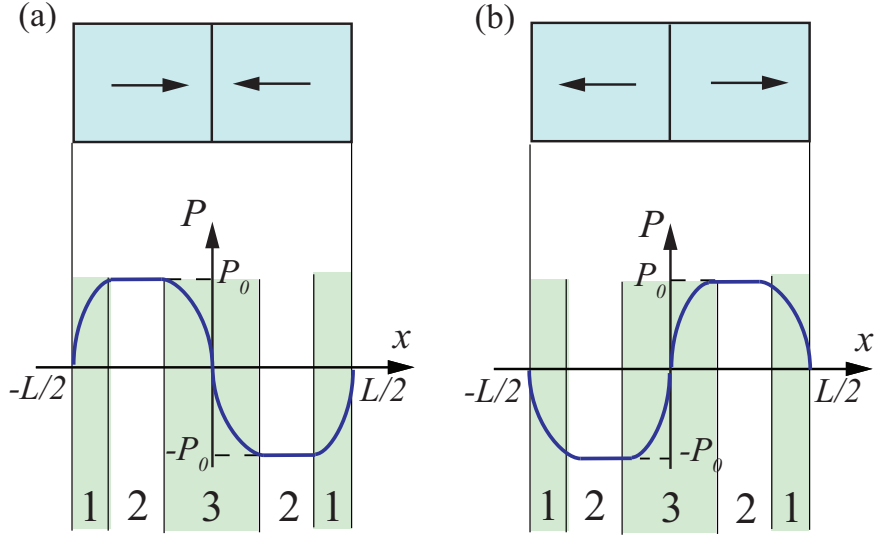}
\caption{Schematic polarization distribution in the case of "head-to-head" (a) and "tail-to-tail" (b) domain walls. Three different regions are marked for each configuration: 1- surface adjacent region, 2 - internal domain region, 3 - domain wall.}
\label{isolated_domains}
\end{center}
\end{figure}

Let us first consider a "head-to-head" domain wall in a large sample, so that the polarization inside the domains is close to the spontaneous polarization.
Thus, neglecting thin surface layer, the polarization distribution can be schematically presented  as shown in Fig. \ref{isolated_domains}.
We define the electrical potential as equal to zero in some point inside the domains (region 2, Fig.  \ref{isolated_domains}),
where the electron concentration is equal to the homogeneous one.
For the "head-to-head" domain wall (region 3, Fig.  \ref{isolated_domains}(a)), for the situation addressed, the problem can be formulated as the search of a solution to the set of equations  \eqref{state}-\eqref{charge_density}  with the boundary conditions
\begin{equation}\label{boundary1}
P|_{-\infty}=P_0, P|_{\infty}=-P_0.
\end{equation}

The polarization distribution in the surface adjacent region (region 1, Fig.\ref{isolated_domains}(a)) can be found from the same equations with the following boundary conditions:
\begin{equation}\label{boundary_surface1}
P|_{-\infty}=-P_0, P|_{L/2}=0,
\end{equation}
\begin{equation}\label{boundary_surface2}
P|_{\infty}=P_0, P|_{-L/2}=0.
\end{equation}

This problem is equivalent to the problem for the "tail-to-tail" domain wall (region 3, Fig.  \ref{isolated_domains}(b)),
 which can be found as the solution to Eq.  \eqref{combined_simp} with boundary conditions:
\begin{equation}\label{boundary2}
P|_{-\infty}=-P_0, P|_{\infty}=P_0.
\end{equation}
Specifically, the polarization profile near the surface of the sample with "head-to-head" wall will be a half of the polarization profile in the "tail-to-tail" domain wall.
In turn, the solution for the "tail-to-tail" domain wall is equivalent to the solutions for the "head-to-head" wall,
 but the screening is provided by holes instead of electrons.
To get the results to the "tail-to-tail" domain wall,
the effective mass and homogeneous concentration of the electrons in the results for the "head-to-head" wall
should be replaced by the mass and homogeneous concentration of holes and vice versa.
Further in this section we will consider the specific case of a "head-to-head" domain wall.

We will consider the limiting cases, when analytical solutions are possible.
For the case $\kappa\gg (\partial \rho/\partial \varphi)^{-1}$ one can rewrite Eq. \eqref{big_scale_combined} as:
\begin{equation}\label{big_scale_r_c}
\alpha P+\beta P^3=\kappa\frac{\partial^2P}{\partial x^2}.
\end{equation}
This is the same equation as well known equation for the neutral domain wall.
The coefficient $\kappa$ is different for the cases when the polarization gradient is parallel and perpendicular to the direction of polarization, but typically this coefficient are of the same order.
The corresponding solution for the "head-to-head" wall is:
\begin{equation}\label{big_scale_solution_r_c}
P=-P_0\tanh\left(\frac{x}{2r_\mathrm{c}}\right).
\end{equation}
In the opposite case $\kappa\ll (\partial \rho/\partial \varphi)^{-1}$,  i.e. $\delta\gg r_\mathrm{c}$ this equation can be simplified as:
\begin{equation}\label{combined_simp}
\frac{\partial \rho}{\partial \varphi}\left(\alpha P+\beta P^3\right)=-\frac{\partial^2P}{\partial x^2}.
\end{equation}

We will address four qualitatively different limit situations, where analytical dependence $\rho(\varphi)$ can be obtained and used in getting the explicit form of $\partial\rho/\partial\varphi$.
First of all, depending on the spontaneous polarization and the homogeneous concentration of electrons, the electron gas in the conduction band inside the domain wall can be degenerate or non-degenerate.
In addition, linear and nonlinear regimes of screening are possible.
We should mention that the gas inside the wall can be degenerate, even if the electron gas in the conduction band in homogeneous ferroelectric is non-degenerate.
We will consider the cases of intrinsic semiconductor or fully ionized doping impurities.

%%%%%%%%%%%%%%  CLASSICAL GAS    %%%%%%%%%%%%%%%%%%%%%%%%%
%%%%%%%%%%%%%%%%%%%%%%%%%%%%%%%%%%%%%%%%%%%%%%%%%

\subsection{Screening with classical electron gas}\label{Classic}

In the case of non-degenerate electron gas which obeys classical statistics, Eq. \eqref{charge_density} can be rewritten in the form:
\begin{equation}\label{concentration:class}
\rho=q\left(-n_\mathrm{e0}e^\frac{q\varphi}{kT}+n_\mathrm{h0}e^{-\frac{q\varphi}{kT}}+n_\mathrm{e0}-n_\mathrm{h0}\right)
\end{equation}
where $n_\mathrm{e0}$ and $n_\mathrm{h0}$ are the homogeneous concentration of electrons and holes, respectively.
This leads directly to:
\begin{equation}\label{concentration:class_diff}
\frac{\partial\rho}{\partial\varphi}=-\frac{q^2}{kT}\left(n_\mathrm{e0}e^\frac{q\varphi}{kT}+n_\mathrm{h0}e^{-\frac{q\varphi}{kT}}\right).
\end{equation}
Together with Eq.\eqref{combined_simp} this leads to:
\begin{equation}\label{combined_class}
\frac{q^2}{kT}\left(n_\mathrm{e0}e^\frac{q\varphi}{kT}+n_\mathrm{h0}e^{-\frac{q\varphi}{kT}}\right)\left(\alpha P+\beta P^3\right)=\frac{\partial^2P}{\partial x^2}.
\end{equation}
Substituting Eq. \eqref{concentration:class} into Eq. \eqref{Poisson_P},  the Poisson equation for classical gas reads:

\begin{equation}\label{poisson_classic}
\frac{\partial P}{\partial x}=-q\left(n_\mathrm{e0}e^\frac{q\varphi}{kT}-n_\mathrm{h0}e^{-\frac{q\varphi}{kT}}-n_\mathrm{e0}+n_\mathrm{h0}\right)
\end{equation}
This is a quadratic equation with respect to $e^\frac{q\varphi}{kT}$. It has only one positive solution:
\begin{equation}\label{classic_exponent}
e^\frac{q\varphi}{kT}=
\frac{
n_\mathrm{e0}-n_\mathrm{h0}-\frac{1}{q}\frac{\partial P}{\partial x}+\sqrt{
\left(n_\mathrm{e0}-n_\mathrm{h0}-\frac{1}{q}\frac{\partial P}{\partial x}\right)^2+4n_\mathrm{e0}n_\mathrm{h0}
}
}
{2n_\mathrm{e0}},
\end{equation}
 and together with Eq.~\eqref{combined_class} it leads to the following equation:
\begin{equation}\label{classic}
\alpha P+\beta P^3=\frac{kT}{q^2}\frac{\partial^2P/\partial x^2}{\sqrt{\left(-\frac{1}{q}\frac{\partial P}{\partial x}-n_\mathrm{e0}+n_\mathrm{h0}\right)^2+4n_\mathrm{h0}n_\mathrm{e0}}}.
\end{equation}
Such equation in the context of the problem of polarization screening was obtained by Guro et al.\cite{Guro68}
In the case of linear screening, the term $\frac{\partial P}{\partial x}$ in Eq.~\eqref{classic} can be neglected in comparison with $n_\mathrm{e0}+n_\mathrm{h0}$ and Eq.~\eqref{classic} can be rewritten in the form:
\begin{equation}\label{classiclargen}
\alpha P+\beta P^3-\frac{kT}{q^2(n_\mathrm{e0}+n_\mathrm{h0})}\frac{\partial^2P}{\partial x^2}=0.
\end{equation}
The exact solution to this equation reads:
\begin{equation}\label{ClasLin:solution}
P=-P_0\tanh \left(\frac{x}{\delta_\mathrm{cl}}\right)
\end{equation}
where
\begin{equation}\label{classicwidth}
\delta_\mathrm{cl}=\sqrt{\frac{2kT}{q^2(n_\mathrm{e0}+n_\mathrm{h0})|\alpha|}}.
\end{equation}
In the case of a doped material, the concentration of minor carriers can be neglected, for example in the case of donor doping, Eq.~\eqref{classicwidth} can be rewritten as:
\begin{equation}\label{classicwidth_e}
\delta_\mathrm{cl}=\sqrt{\frac{2kT}{q^2n_\mathrm{e0}|\alpha|}}
\end{equation}
This spatial scale is about the Debye screening length for a linear media with a permittivity about $\varepsilon_\mathrm{f}\approx2\pi/|\alpha|$.
If the concentration of electrons inside the domain wall is much larger than the homogeneous electron and hole concentrations inside the domain, Eq.~\eqref{classic} can be rewritten in the form:
\begin{equation}\label{classicNL}
\alpha P+\beta P^3=-\frac{kT}{q}\frac{\partial^2P/\partial x^2}{{\partial P}/{\partial x}}.
\end{equation}

Changing variables $P=P_0p$, $x=\delta_\mathrm{cl}^\mathrm{nl}y$, where:
\begin{equation}\label{classic_nonlinear_width}
\delta_\mathrm{cl}^\mathrm{nl}=\frac{2kT}{q|\alpha|P_0}.
\end{equation}
Equation~\eqref{classicNL} can be presented in a dimensionless form:
\begin{equation}\label{ClasNL:dimless}
\left(p-p^3\right)\frac{\partial p}{\partial y}=\frac{1}{2}\frac{\partial^2p}{\partial y^2}.
\end{equation}
The same characteristic scale $\delta_\mathrm{cl}^\mathrm{nl}$ was obtained by Krapivin and Chenskii\cite{Chenskii}.
It corresponds to the nonlinear Debye screening length in linear media with permittivity about $2\pi/|\alpha|$, which is the typical length of charge screening in the case of classical statistics when homogeneous electron concentration is small in comparison with electron concentration in the screening area. The exact solution to this equation with boundary conditions $p(-\infty)=1$, $p(\infty)=-1$, can be written as:
\begin{equation}\label{ClasNL:exsol}
y=-\frac{p}{4(1-p^2)}+\frac{1}{8}\ln\left(\frac{1-p}{1+p}\right).
\end{equation}

\begin{figure}
\begin{center}
\includegraphics[width=0.5\textwidth]{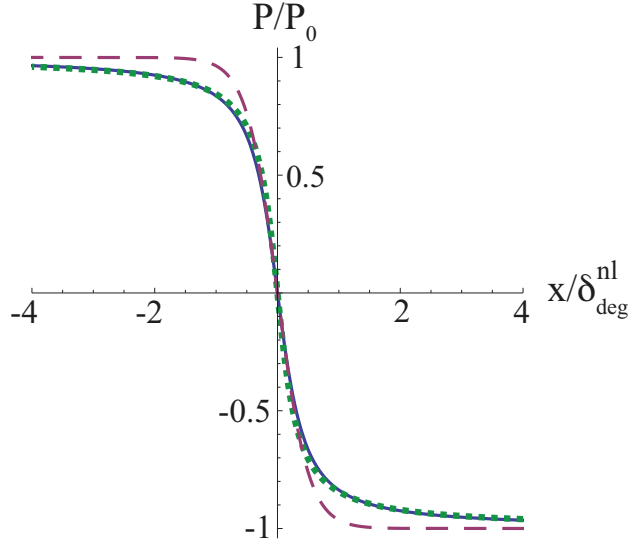}
\caption{Exact polarization profile Eq. \eqref{ClasNL:exsol} (solid line), approximation $p=-\tanh 2y$ (dashed line) and approximation $p=-(2/\pi)\arctan 4y$ (dotted line).}
\label{numsol_cl}
\end{center}
\end{figure}

 The solution is presented in Fig.~\ref{numsol_cl}. It can be approximated with the function $p=-(2/\pi)\arctan 4y$. The function $p=-\tanh 2y$ is given for comparison.

This solution describes properly only the part of domain wall, where concentration is much larger than the homogeneous concentration.
Thus the polarization profile given by Eq. \eqref{ClasNL:exsol} applicable to the case where the screening regime is nonlinear in most part of the domain wall.

%%%%%%%%%%%%%%%%%%%%%%%%%DEGENERATE
%%%%%%%%%%%%%%%%%%%%%%%%%%%%%%%%%%%%%%%%%%%%%%

\subsection{Screening with degenerate electron gas}\label{Degenerate}

The density of states in valence and conduction bands for degenerate gas with parabolic spectrum can be obtained from the effective electron and hole masses as follows \cite{Sze}:
 \begin{equation}\label{DOS_cond}
N_\mathrm{C}=2\left(\frac{m_\mathrm{e}kT}{2\pi\hbar^2}\right)^{3/2}
\end{equation}
 \begin{equation}\label{DOS_val}
N_\mathrm{V}=2\left(\frac{m_\mathrm{h}kT}{2\pi\hbar^2}\right)^{3/2}
\end{equation}
where $m_\mathrm{e}$ is the effective electron mass and $m_\mathrm{h}$ is the effective hole mass. Below we neglect the difference in the densities of state in the valence and the conduction bands and between electron and hole effective masses.
For numerical calculations we use the free electron mass $m$.

For a degenerate gas, the Fermi functions can be approximated by a step function and Dirac-Fermi integral in Eq.~\eqref{charge_density} can be presented as
 \begin{equation}\label{DF_integral}
F_\mathrm{1/2}(z)=\frac{4}{3\sqrt{\pi}}z^{3/2}
\end{equation}
Two cases are possible in this approximation.
Either valence band is fully occupied and there are electrons in conduction band, or there are holes in the valence band and the conduction band is empty.
 In the first case, using Eqs. \eqref{DOS_cond}, \eqref{DF_integral}, Eq.~\eqref{charge_density} can be rewritten as:
\begin{equation}\label{n_degen1}
\rho=-q\left(\frac{(2m(q\varphi+E_\mathrm{F}-E_\mathrm{C}))^{3/2}}{3\pi^2\hbar^3}-n_\mathrm{e0}\right).
\end{equation}

In the second case, using Eqs. \eqref{DOS_val}, \eqref{DF_integral}, Eq.~\eqref{charge_density} the charge density can be found as follows:
\begin{equation}\label{n_degen2}
\rho=q\left(\frac{(2m(-q\varphi+E_\mathrm{V}-E_\mathrm{F}))^{3/2}}{3\pi^2\hbar^3}-n_\mathrm{h0}\right).
\end{equation}
Both Eqs. \eqref{n_degen1} and \eqref{n_degen2} lead to the following result for $\partial\rho/\partial\varphi$:
\begin{equation}\label{drhodphi_degen}
\frac{\partial\rho}{\partial\varphi}=-\frac{3mq^2}{(3\pi^2)^{2/3}\hbar^2}\left(\frac{|\rho|}{q}+n_0\right)^{1/3}.
\end{equation}
where $n_0$ is the homogeneous concentration of the majority carriers.

Together with  \eqref{combined_simp} and \eqref{Poisson_P} this leads to an equation for the polarization inside the domain wall in the following form:
\begin{equation}\label{degen_gen}
\frac{3mq^2}{(3\pi^2)^{2/3}\hbar^2}\left(\frac{1}{q}\left|\frac{\partial P}{\partial x}\right|+n_0\right)^{1/3}\left(\alpha P+\beta P^3\right)=\frac{\partial^2 P}{\partial x^2}.
\end{equation}

In the case of linear screening, when the concentration of electrons is just slightly changed in comparison with their homogeneous concentration in the conduction band or the concentration of holes is slightly changed in comparison with homogeneous holes concentration in the valence band, Eq.~\eqref{degen_gen}   can be rewritten in the form:
\begin{equation}\label{degen_lin}
\alpha P+\beta P^3-\frac{(3\pi^2)^{2/3}\hbar^2}{3mq^2n_0^{1/3}}\frac{\partial^2P}{\partial x^2}=0
\end{equation}
This equation can be solved analytically, with the boundary conditions $P(-\infty)=P_0$, $P(\infty)=-P_0$, like in the case of uncharged wall, and the polarization can be found as follows:
\begin{equation}\label{exactPdeg}
P=-P_0\tanh \left(\frac{x}{\delta_\mathrm{deg}}\right)
\end{equation}
where
\begin{equation}\label{delta:deg}
\delta_\mathrm{deg} = \sqrt{\frac{2(3\pi ^2)^{2/3}\hbar ^2}{3mq^2n_0^{1/3}|\alpha|}}
\end{equation}
is the typical half-width of a domain wall. This scale is the Thomas-Fermi screening length in linear media where the permittivity $\varepsilon$ is about  $2\pi/|\alpha|$:
\begin{equation}\label{TF}
\lambda_\mathrm{TF} =\frac{1}{2}\sqrt{\frac{\pi^{1/3}\hbar ^2\varepsilon}{3^{1/3}mq^2n_0^{1/3}}}.
\end{equation}

In the case of nonlinear screening, where the electron concentration in the conduction band inside the wall is much larger then the homogeneous concentration, Eq.~\eqref{degen_gen}  can be rewritten in the form:
\begin{equation}\label{degen_nonlinear}
\alpha P+\beta P^3+\frac{(3\pi^2)^{2/3}\hbar^2}{3mq^{5/3}}\frac{\partial^2P/\partial x^2}{(\partial P/\partial x)^{1/3}}=0.
\end{equation}

Changing variables:
\begin{equation}\label{Pdimless}
 P=P_0p,
\end{equation}
\begin{equation}\label{xdimless}
 x=\delta_\mathrm{deg}^\mathrm{nl}y,
\end{equation}
where
\begin{equation}\label{delta:degnl}
\delta_\mathrm{deg}^\mathrm{nl} = \left(\frac{9\pi^4\hbar ^6}{q^5m^3|\alpha|^3P_0}\right)^{1/5},
\end{equation}
Eq.~\eqref{degen_nonlinear} can be transformed to a dimensionless form:
\begin{equation}\label{degen_nonlinear_dimensionless}
p-p^3-\frac{1}{3}\frac{\partial^2 p/\partial y^2}{\left(\partial p/\partial y\right)^{1/3}}=0.
\end{equation}
Thus, the characteristic scale for the domain wall half-width in this case is about $\delta_\mathrm{deg}^\mathrm{nl}$,
which is of the order of the characteristic scale for nonlinear Thomas-Fermi screening for linear dielectrics with a permittivity of about $2\pi/|\alpha|$,
i.e. the typical screening scale in the case where the electron gas in screening area is degenerate with the concentration much larger than the homogeneous concentration.
The same scale (to within a factor of the order of unity) was obtained by Ivanchik\cite{Ivanchik} for the case of internal screening in the monodomain state in a ferroelectric which is an intrinsic semiconductor.

\begin{figure}
\begin{center}
\includegraphics[width=0.5\textwidth]{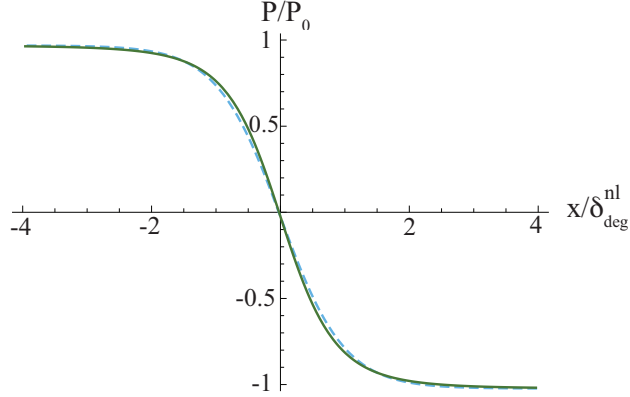}
\caption{Exact polarization profile Eq. \eqref{degen_nonlinear_exact} (solid line) and approximation $p=-\tanh y$ (dashed line).}
\label{numsol}
\end{center}
\end{figure}

The exact solution of Eq. \eqref{degen_nonlinear_dimensionless}  can be obtained in the following form:
\begin{equation}\label{degen_nonlinear_exact}
y=-\left(\frac{4}{5}\right)^{3/5}\int_0^p\frac{1}{(1-\xi^2)^{6/5}}d\xi
\end{equation}
The solution is presented in Fig.~\ref{numsol}. It can be approximated with the function $p=-\tanh y$.

Similar to the case of nonlinear screening with classical electron gas, this obtained polarization profile applies to the case, where the screening regime is nonlinear in the most part of the domain wall.
It is clear that the solution obtained above can be used also in the case of non-degenerate electron gas in the bulk of the domains, whereas inside the domain wall the gas is degenerate.

%%%%%%%%%%%%%%%%%%%%%%%%%%%%%%%%%%%
%%%%%%%COMPARISON

\subsection{Charged domain wall in a large sample. Applicability for different cases.}\label{Comparison}

\begin{figure}[h]
\begin{center}
\includegraphics[width=0.5\textwidth]{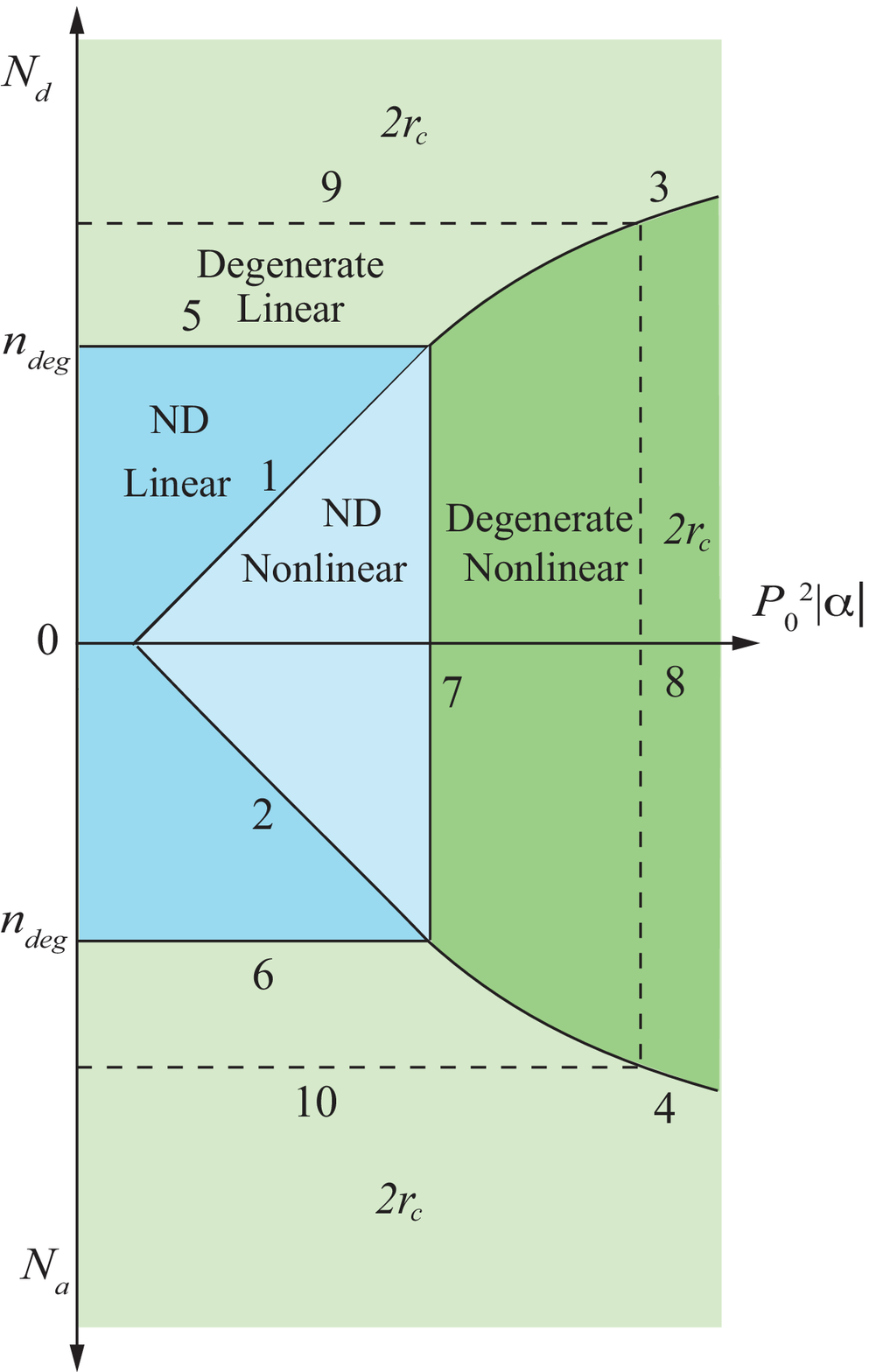}
\caption{A diagram presenting areas that correspond to the different regimes of screening in a "head-to-head" domain-wall in a large ferroelectric crystal, depending on the ferroelectric energy density $|\alpha|P_0^2$  and the concentration of the fully ionized dopant. Here $N_\mathrm{d}$  and  $N_\mathrm{a}$ are the donor and acceptor dopant concentrations, respectively. Inside the region separated by dashed line, the domain-wall half-width corresponds to one of the scales obtained in Sect. \ref{Analytical_Solutions}. Outside this region the  gradient term becomes important and the  domain-wall half-width is about the correlation radius $r_c$.}
\label{diagram2}
\end{center}
\end{figure}

Now, that the domain wall width for all regimes of screening is known, we can determine the conditions of applicability for each case.

The electron concentration in the center of the "head-to-head" domain wall can be estimated as:
\begin{equation}\label{concentration_wall}
n_\mathrm{c}=\frac{P_0}{q\delta}+n_0.
\end{equation}
where $\delta$ is the domain wall half-width.

There are three main conditions, which subdivide the different regimes for domain wall screening. The condition of applicability of the formula for a strongly degenerate Fermi gas requires a small temperature in comparison to the Fermi temperature,\cite{Landau5} i.e.
\begin{equation}\label{condition_degen}
kT \ll \frac{(3\pi^2)^{2/3}\hbar ^2}{2m}n_\mathrm{c}^{2/3}.
\end{equation}
In the opposite case the gas is classical.

The condition for the linear regime of screening has the form:
\begin{equation}\label{condition_linear}
n-n_0=P_0/q\delta\ll n_0,
\end{equation}
hereafter $n_0=\mathrm{max}(n_\mathrm{e}, n_\mathrm{h})$.

If the obtained half-width
\begin{equation}\label{condition_rc}
\delta<\delta_\mathrm{neutral},
\end{equation}
the gradient term plays a decisive role and the domain wall half-width will be about the $\delta_\mathrm{neutral}=2r_c$.

The diagram presented in Fig.~\ref{diagram2} was obtained with conditions Eqs. \eqref{condition_degen}-\eqref{condition_rc}.
It is instructive to plot it in the coordinates of full-ionized dopants concentration instead of $n_0$.
The zero doping corresponds to $n_0=2n_\mathrm{i}$, where $n_\mathrm{i}$ is the intrinsic carriers concentration.

For a non-degenerate gas, the line separating linear and nonlinear screening (lines 1, 2 in Fig.~\ref{diagram2}) can be described by the equation:
\begin{equation}\label{lim:nondeg}
n_0=\frac{|\alpha|P_0^2}{kT}.
\end{equation}

For the degenerate gas, the regions corresponding to the linear and nonlinear screening are separated by the lines  3 and 4 in Fig.~\ref{diagram2} and can be described by the equation:
\begin{equation}\label{lim:deg}
n_0=\left(\frac{|\alpha|P_0^2m}{\hbar^2}\right)^{3/5}.
\end{equation}

Lines 5, 6 and 7 in Fig.~\ref{diagram2} separate the cases of screening by degenerate and classical gases and can be described by equations
\begin{equation}\label{lim:degLin}
n_0=n_\mathrm{deg}=\frac{1}{3\pi^2}\left(\frac{2kTm}{\hbar^2}\right)^{3/2}
\end{equation}
and
\begin{equation}\label{lim:degNl}
|\alpha|P_0^2=\frac{1}{3\pi^2}\frac{\left( 2kT\right)^{5/2}m^{3/2}}{\hbar^3},
\end{equation}
respectively.

Lines 8, 9, and 10 separate the regions where the gradient term is important and can be described by equations :
\begin{equation}\label{lim:nrc}
n_0=\left(\frac{(3\pi^2)^{2/3}\hbar^2}{3mq^2\kappa}\right)^3=\frac{\pi^4}{3}\left(\frac{a_0}{\kappa}\right)^3.
\end{equation}
\begin{equation}\label{lim:Prc}
|\alpha|P_0^2=\frac{(3\pi^2)^4}{2^5}\frac{\hbar^{12}}{m^6q^{10}\kappa^5}=\frac{(9\pi^4a_0^3q)^2}{(2\kappa)^5},
\end{equation}
 where $a_0=\hbar^2/mq^2$ is the Bohr radius.

The regimes of screening presented in this diagram work in extreme cases far from the borders of the corresponding regions.
Near the borders, this regimes of screening will show a crossover behavior.
Lines 2 and 4 on this diagram correspond to the full depletion.
On these lines, the characteristic scales for linear and nonlinear screening can be matched, and the domain-wall half-width will be about the depletion width for the holes:
\begin{equation}\label{depletion}
\delta_\mathrm{dep}=\frac{P_0}{qn_\mathrm{a}},
\end{equation}

For the case of typical perovskites like BaTiO$_3$ and PbTiO$_3$ at room temperature, the domain wall corresponds to the regime of nonlinear screening by degenerate gas (see Sect. \ref{Experiment}).

 %%%%%%%%%%%%%%%%%%%%%%%%%%%

\section{Formation energy of a charged domain wall}\label{Energies}

Now, that the polarization profile is known, we can calculate the energy per unit area of the isolated sample with charged domain wall, defined by Eq. \eqref{DWenergy}.
First we will neglect the energy associated with the surface boundary layer with atomic thickness.
In this case the energy of the sample with charged domain wall are the same for "head-to-head" and "tail-to-tail" configurations.
 If the sample is large enough, the impact on the energy of the states with the domain-wall can be calculated separately for the domain-wall and the surface adjacent regions of the sample
 for both configurations shown in Fig. \ref{isolated_domains}.
Then, the formation energy of the charged domain wall, Eq. \eqref{DWenergy}, can be rewritten as follows:
\begin{equation}\label{W_separate}
W=\int_{-\infty}^{\infty}[ \Phi_\mathrm{h-h}(x)-\Phi_0(x)]dx+\int_{-\infty}^{\infty}[ \Phi_\mathrm{t-t}(x)-\Phi_0(x)]dx.
\end{equation}
where $\Phi_\mathrm{h-h}(x)$ and $\Phi_\mathrm{t-t}(x)$ are the free energies per unit area for the polarization profiles calculated with boundary conditions \eqref{boundary1} and \eqref{boundary2}, respectively.

Using Eq. \eqref{state}, the free energy density, Eq. \eqref{fe}, can be rewritten as:
\begin{equation}\label{fe_state}
\Phi=\frac{\alpha}{2}P^2+\frac{\beta}{4}P^4 + \frac{\kappa}{2}\left(\frac{\partial P}{\partial x}\right)^2 + \frac{\varepsilon_\mathrm{b}}{8\pi}\left(\alpha P+\beta P^3 - \kappa\left(\frac{\partial^2P}{\partial x^2}\right)\right)^2+\Phi_\mathrm{eg}.
\end{equation}
We can neglect the electrostatic contribution in free energy in comparison with the ferroelectric one with respect to the small parameter $\varepsilon_\mathrm{b}/\varepsilon_\mathrm{f}$.
In our problem $|P|\leq|P_0|$ that leads to $|\beta P^3|\leq|\alpha P|$, $|\kappa(\partial^2 P/\partial x^2)|\leq|\kappa P/r_\mathrm{c}^2|=|2\alpha P|$, and electrostatic term can be estimated as:
\begin{equation}\label{fe_electrost}
\left| \frac{\varepsilon_\mathrm{b}}{8\pi}\left(\alpha P+\beta P^3 - \kappa\left(\frac{\partial^2P}{\partial x^2}\right)\right)^2\right|\leq\frac{\varepsilon_\mathrm{b}}{2\pi}|\alpha|^2P^2
=\frac{\varepsilon_\mathrm{b}}{\varepsilon_\mathrm{f}}|\alpha|P^2,
\end{equation}
 and we can neglect it in comparison with $\alpha P^2$.
 Using $\delta\gg r_\mathrm{c}$ we can also neglect gradient term $\frac{\kappa}{2}\left(\frac{\partial P}{\partial x}\right)^2$ in comparison with $\alpha P^2$.
 Finally, Eq. \eqref{fe_state} can be rewritten as:
 \begin{equation}\label{fe_simp}
\Phi=\frac{\alpha}{2}P^2+\frac{\beta}{4}P^4+\Phi_\mathrm{eg}.
\end{equation}

For the case of a classical gas, the free energy density may be written as (see App. \ref{appendix1}):
\begin{equation}\label{energy_classicaleg}
\Phi(x)-\Phi_0=\left(\frac{\alpha}{2}P^2+\frac{\beta}{4}P^4-\frac{\alpha}{4}P_0^2\right)+\left(q\varphi(n_\mathrm{e}-n_\mathrm{h})-kT(n_\mathrm{e}+n_\mathrm{h}-n_\mathrm{e0}-n_\mathrm{h0})\right).
\end{equation}
Here, the first term corresponds to the lattice energy and the second one to the free kinetic energy of the electron gas.

As shown in App. \ref{appendix2}, the energy of the charged wall linearly screened by classical electron gas can be obtained
from Eqs. \eqref{W_separate}, \eqref{energy_classicaleg} using solution Eq. \eqref{ClasLin:solution} in the form:
\begin{equation}\label{W_cl_lin}
W_\mathrm{cl}=\frac{4}{3}|\alpha|P_0^2\delta_\mathrm{cl}.
\end{equation}

For the case of nonlinear screening by classical gas, using Eqs.\eqref{classicNL} and \eqref{energy_classicaleg}, the charged domain-wall energy can be presented in the form (see App. \ref{appendix2}):
\begin{equation}\label{W_cl_nl2}
W_\mathrm{cl}^\mathrm{nl}=\frac{2P_0}{q}\left(E_\mathrm{g}+2kT\ln\left(\frac{4|\alpha|P_0^2}{kTN}-8kT\right)\right),
\end{equation}
where $N$ is the density of states in the valence and the conduction bands.
It can be shown (see App. \ref{appendix2}) that for $E_\mathrm{g}\gg kT $, the second term in the parenthesis can be neglected in comparison with the first one, and the $W_\mathrm{cl}^\mathrm{nl}$ is given by:
\begin{equation}\label{W_cl_nl}
W_\mathrm{cl}^\mathrm{nl}=\frac{2P_0}{q}E_\mathrm{g}.
\end{equation}

For the case of a degenerate gas, the free energy density can be presented in the following form (see Appendix \ref{appendix1}):
\begin{equation}\label{energy_Degenerate_eg}
\Phi(x)=\left(\frac{\alpha P^2}{2}+\frac{\beta P^4}{4}-\frac{\alpha P_0^2}{4}\right)+\left((n_\mathrm{e}-n_\mathrm{e0})E_\mathrm{g}+\frac{3\hbar ^2(3\pi ^2)^{2/3}}{10m}(n_\mathrm{e}^{5/3}+n_\mathrm{h}^{5/3}-n_\mathrm{e0}^{5/3}-n_\mathrm{h0}^{5/3})\right).
\end{equation}

As shown in Appendix \ref{appendix2}, the charged domain wall energy in the case of linear screening by degenerate gas can be expressed in the form:
\begin{equation}\label{W_deg_lin}
W_\mathrm{deg}=\frac{4}{3}|\alpha|P_0^2\delta_\mathrm{deg},
\end{equation}
and in the case of nonlinear screening
\begin{equation}\label{W_deg_nl}
W_\mathrm{deg}^\mathrm{nl}=0.77|\alpha|P_0^2\delta_\mathrm{deg}^\mathrm{nl}+\frac{2P_0}{q}E_\mathrm{g}.
\end{equation}

It can be shown (see Appendix \ref{appendix2}) that for realistic values of spontaneous polarization and band gap (for $E_\mathrm{g}=3$ eV, $P_0<300$ $\mathrm{\mu C/cm}^2$), the first term in Eq. \eqref{W_deg_nl} can be neglected in comparison with the second one, and the energy reads:
\begin{equation}\label{W_deg_nl1}
W_\mathrm{deg}^\mathrm{nl}=\frac{2P_0}{q}E_\mathrm{g}.
\end{equation}
This  relation is universal in the case of nonlinear screening for degenerate and classical gas.
This is the energy needed in order to create sufficient electron-hole pairs for polarization screening.
Although this result was strictly obtained for the case of a second order phase transition,
one can show it is also valid for materials with a first order phase transition.
One should note that if we neglect the surface energy the energies of configurations with charged domain wall are the same for "head-to-head" and "tail-to-tail" walls (see Fig. \ref{isolated_domains}).

In the frame of the continuous theory the energy of the thin boundary layer can be taken into account introducing a phenomenological term $2\zeta_1\vec{P_\mathrm{d}}\vec{n}$ in the formation energy of the domain wall, where $\vec{P_\mathrm{d}}$ is the polarization inside the domain, $\vec{n}$ is the normal to the surface directed toward the sample and $\zeta_1$ is the effective coefficient in the surface energy (here we restricted our consideration with linear term only).
To obtain this coefficient the microscopical consideration is required.
The difference between the energy of "head-to-head" and "tail-to-tail" domain wall is $4\zeta_1\vec{P_0}\vec{n}$.
In the case of nonlinear screening the surface energy can be taken into account, introducing the effective changing of the bandgap:
\begin{equation}\label{E_eff}
E_\mathrm{g}^\mathrm{eff}=E_\mathrm{g}\pm q\zeta_1,
\end{equation}
for "head-to-head" and "tail-to-tail" walls respectively.
The formation energy of charged domain wall with such notation have the form:
\begin{equation}\label{W_nl_eff}
W^\mathrm{nl}=\frac{2P_0}{q}E_\mathrm{g}^\mathrm{eff}
\end{equation}

\section{Size effect}\label{Size}

\begin{figure}[h]
\begin{center}
\includegraphics[width=0.6\textwidth]{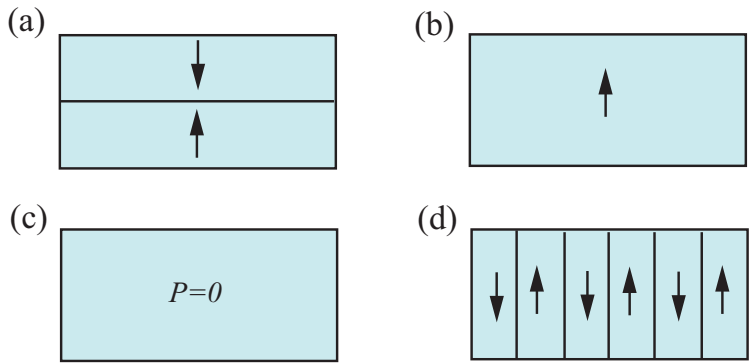}
\caption{Schematics of polarization distribution for a state with charged wall (a), monodomain state (b), state with zero polarization (c), and the multidomain state with antiparallel domains separated by neutral domain walls (d).}
\label{Energies_size}
\end{center}
\end{figure}

We will now address the question of the conditions for which the "head-to-head" configuration is energetically favorable.
Three other possible configurations: singledomain state, configuration with polarization equal to zero and multidomain state with antiparallel domains separated by neutral domain walls (see Fig. \ref{Energies_size}) will be considered as competing scenarios.
Hereafter we will be considering the case of practical importance where the charged domain wall is in the nonlinear screening regime.
Let us start from the competition between the "head-to-head" ("tail-to-tail") configuration and the paraelectric state by comparing the corresponding energies.
We will take into account that for the sample with finite thickness $L$ the polarization can be different from the spontaneous one, due to depolarizing effect.
While Eq. \eqref{W_nl_eff} was obtained for the polarization equal the spontaneous one inside the domain, it can be modified for the case of arbitrary polarization inside the domain (if the screening regime in this case is also nonlinear).
In this case the ferroelectric energy inside the domain should be taken into account.
So that, for the sample much thicker than the domain wall width formation energy of charged domain wall can be estimated as:
\begin{equation}\label{fe_finite_domains}
W=L\left(\frac{\alpha}{2}P_\mathrm{d}^2+\frac{\beta}{4}P_\mathrm{d}^4-\frac{\alpha}{4}P_0^2\right)+\frac{2|P_\mathrm{d}|}{q}E_\mathrm{g}^\mathrm{eff},
\end{equation}
where $P_\mathrm{d}$ is the polarization inside the domain.
(Note that discussing the energy of a sample we apply the term energy to the difference between the energy of the sample and it energy of the material in the  singledomain ferroelectric state times the volume of the sample.)
It can be found as a polarization corresponding to the minimum of the energy Eq. \eqref{fe_finite_domains}.
If $E_\mathrm{g}^\mathrm{eff}>0$, the polarization in the sample with charged domain wall $P_\mathrm{d}<P_0$.
If $E_\mathrm{g}^\mathrm{eff}<0$, that is possible either for "tail-to-tail" or "head-to-head" configuration at high polarization effect of the surface,  $P_\mathrm{d}>P_0$.
In a sample, thinner than some critical thickness we will find below, ferroelectricity can be completely suppressed due to the depolarizing field, and the configuration of zero polarization is energetically favorable.
The energy (in the meaning specified above) of this way formed paraelectric state reads
\begin{equation}\label{paraelectric}
W=-L\frac{\alpha}{4}P_0^2
\end{equation}
Thus, from Eqs. \eqref{DWenergy} and \eqref{paraelectric}, the energy difference per unit area between the state with charged domain wall and paraelectric state can be written as:
\begin{equation}\label{W_par}
\Delta W=L\left(\frac{\alpha}{2}P_\mathrm{d}^2+\frac{\beta}{4}P_\mathrm{d}^4\right)+\frac{2|P_\mathrm{d}|}{q}E_\mathrm{g}^\mathrm{eff}.
\end{equation}
The state with charged domain wall is energetically favorable if for some polarization $P_\mathrm{d}$ the difference $\Delta W<0$.
This inequality has a solution if the sample is thicker than the critical value $L_\mathrm{par}$ defined as:
\begin{equation}\label{fe_der_para_vs_domains2}
L_\mathrm{par}=\frac{3\sqrt{6}E_\mathrm{g}^\mathrm{eff}}{q|\alpha|P_0}\approx\frac{7.3\times E_\mathrm{g}^\mathrm{eff}}{q|\alpha|P_0}.
\end{equation}
Thus, for sample thickness larger than $L_\mathrm{par}$ the "head-to-head" configuration is more energetically favorable than the paraelectric state, otherwise the latter is favorable.

It is worth noting such critical thickness is $\sqrt{2}$ times larger than that that can be obtained from the condition 
\begin{equation}\label{crit}
E_\mathrm{th}> E_\mathrm{g}^\mathrm{eff}/(qL), 
\end{equation}
where $E_\mathrm{th}$ is the thermodynamic coercive electric field. 
Such condition, written with the neglect of the surface poling effect was offered by Ivanchik \cite{Ivanchik} for the stability of a ferroelectric state with internal screening.

Next we discuss the competition between the  "head-to-head" ("tail-to-tail") configuration and the internally screened single-domain state.
In the singledomain configuration the linear terms in surface energy is cancelled out and the energy of the internally screened singledomain state equals a half of the energy of the configuration with charged wall, without surface energy term.
In the case of nonlinear screening, one can find it as follows:
\begin{equation}\label{W_mon}
W_\mathrm{mon}=\frac{P_0}{q}E_\mathrm{g}
\end{equation}
(In writing this equation and further in the paper we  cover only  the situation in samples with the thickness $L\gg L_\mathrm{par}$ where we can neglect difference between $P_\mathrm{d}$ and $P_0$, thus we will use the energy given by Eq. \eqref{W_nl_eff}.)
This energy has the same meaning as Eq. \eqref{W_deg_nl1}, i.e. this is the energy that is necessary for the creation of sufficient electron-hole pairs for the polarization screening.
Here an important remark should be made.
In their treatment of the problem of monodomain state screening, Ivanchik\cite{Ivanchik} and Watanabe\cite{Watanabe} did not take into account the energy needed for the creation of electron-hole pairs.
Therefore they obtained a much smaller energy for the internally screened monodomain state,
leading to a much milder condition for the internally screened domains to be favorable in comparison with the domain screened by the charge in the electrodes.

\begin{figure}[h]
\begin{center}
\includegraphics[width=0.6\textwidth]{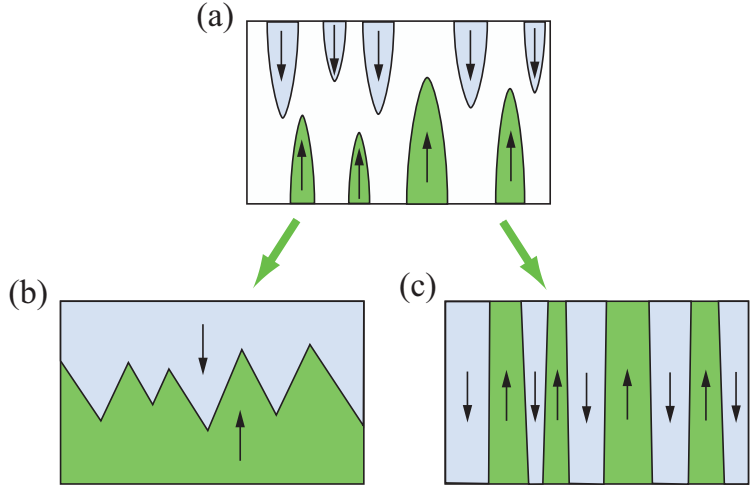}
\caption{Schematics of possible domains growth processes. Domains growing in the form of elongated precursors (a), and two possible results of such growing: "head-to-head" domains (b) and multidomain state with neutral domain walls (c).}
\label{Growth}
\end{center}
\end{figure}

Comparing Eqs. \eqref{W_mon} and \eqref{W_nl_eff} one can find that one of the configurations with charged domain wall is energetically favorable if $2E_\mathrm{g}^\mathrm{eff}<E_\mathrm{g}$, i.e. $|\zeta_1|>E_\mathrm{g}$.
if the surface polarizing effect is not big enough, energy of the monodomain state is always smaller than the energy of the state with charged walls,
thus the configuration with a charged wall is metastable, and charged walls might occur only as a result of growth kinetic of domains with elongated precursors (Fig. \ref{Growth}).
As a result of such growth either charged domain wall (Fig. \ref{Growth}b) or lamella pattern with neutral domain walls (Fig. \ref{Growth}c) can be formed.
Now let us discuss the competition between the  "head-to-head" ("tail-to-tail") configuration and lamella pattern with neutral domain walls.
In the case of lamella pattern with neutral domain walls the surface energy contributions are cancels out.
The energy of the sample per unit area for this configuration is proportional to $\sqrt{L}$ and for both first and second order phase transitions it reads\cite{Tagantsev_book} :
\begin{equation}\label{energy_multidomain}
W_\mathrm{d} = 2 \left(\frac {4 \pi \times0.26 \times W _\mathrm{neutral}} {\sqrt{\varepsilon_\mathrm{a} \varepsilon_\mathrm{c}}}\right)^{1/2} P_0\sqrt{L},
\end{equation}
where $\varepsilon_\mathrm{a}$ and $\varepsilon_\mathrm{c}$ are the lattice permittivity measured perpendicular and along the polarization, respectively, and $W _\mathrm{neutral}$ is the energy per unit area of the neutral domain wall.
Comparing energies given in Eqs. \eqref{energy_multidomain} and \eqref{W_nl_eff}, for $E_\mathrm{g}^\mathrm{eff}>0$
one can find that the multidomain state with neutral walls is more energetically favorable that the "head-to-head" configuration if $L$ is smaller than:
\begin{equation}\label{condition_multi}
L_\mathrm{md}=\left(\frac{E_\mathrm{g}^\mathrm{eff}}{q}\right)^2\frac{\sqrt{\varepsilon_\mathrm{a} \varepsilon_\mathrm{c}}}{4 \pi \times0.26  \times W _\mathrm{neutral}}.
\end{equation}

For a second order phase transition and near the transition temperature $T_\mathrm{C}$, the neutral-wall surface-energy tends to zero according to the laws\cite{Levanyuk} $W _\mathrm{neutral}\propto (T_\mathrm{C}-T)^{3/2}$ and $\varepsilon_\mathrm{c}\propto  (T_\mathrm{C}-T)^{-1}$, $\varepsilon_\mathrm{a}$ is constant for uniaxial ferroelectrics and $\varepsilon_\mathrm{a}\propto  (T_\mathrm{C}-T)^{-1}$ for cubic ferroelectrics \cite{Levanyuk}.
Thus $L_\mathrm{md}$ tends to infinity and the multidomain state with neutral walls is always preferable.
The situation is different for materials with first order phase transition, where close to the transition temperature, $L_\mathrm{md}$ is finite.
Here, depending on the sample thickness, either "head-to-head" domains or multidomain state with neutral walls is energetically preferable.
Once one of these configurations is formed passing the phase transition, the spontaneous switching between them, which is related with charge transfer, does not look very probable.

%%%%%%%%%%%%%%%%%%%%%%%%%%%%%%%%%%%%%%%%%%%%%%%

\section{Isolated sample vs. sample with electrodes}\label{electrodes}

The problem of the charged domain wall creation in the presence of metal electrodes was considered in a series of work by Ivanchik, Guro, Vul, and Kovtonyuk\cite{Guro70,Vul} and by Krapivin and Chenskii\cite{Chenskii}.
After the mentioned authors, we will first neglect the surface energy effects, and will consider it separately.
Below we show how some estimates for a sample with electrodes can be obtained, based on the results of the charged wall in the isolated sample.
We will analyse the difference between these cases.

We discuss the practically important case of nonlinear screening.
In this case the energy of the charged domain wall is equal to the energy, necessary to create enough electrons and holes for screening.
In the isolated sample, the only source of screening electrons is the valence band, and the creation of each electron-hole pair increases the free energy by $E_\mathrm{g}$.
In a sample with metal electrodes, a charge carriers exchange between the ferroelectric and the metal is possible, and the metal can serve as the source of the screening electrons or holes.
The charge that will be created at the electrodes due to the electron transfer will screen the bound charge near the surface.

To find the energy necessary for electron transfer we should take into consideration the chemical potential $\mu$ and the work function difference between the ferroelectric and the electrodes.
For each electron that comes from the metal, the energy $E_\mathrm{C}-\mu+A_\mathrm{e}-A_\mathrm{f}$,
where $A_\mathrm{e}$ and $A_\mathrm{f}$ are the work functions for the electrode and ferroelectric respectively,
should be added to the free energy of the system.
The chemical potential for electrons is the Fermi level, and the energy of "head-to-head" domain wall can be estimated as:
\begin{equation}\label{electrodes1}
W=\frac{2P_0}{q}(E_\mathrm{C}-E_\mathrm{F}+A_\mathrm{e}-A_\mathrm{f}).
\end{equation}

\begin{figure}[h]
\begin{center}
\includegraphics[width=0.6\textwidth]{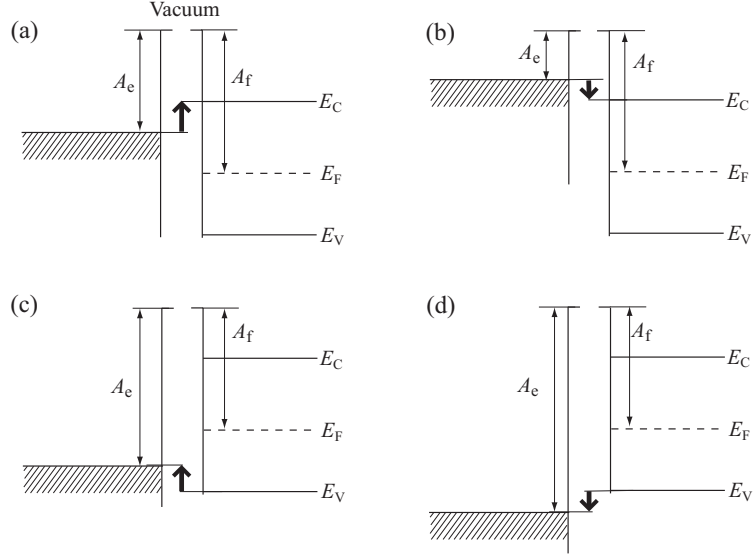}
\caption{Band diagrams for different work-function differences between a ferroelectric and a metal when ferroelectric and metal are not in a physical contact.
When the electrode and the ferroelectric are separated, the shift between the Fermi levels of the materials is defined by the work-function difference.
The energy necessary for the creation of a unit free charge is shown with the bold arrow.
This energy can be negative for "head-to-head" (b), or "tail-to-tail" configuration (d), and the domain wall is energetically preferable in this case.}
\label{Electrodes_band}
\end{center}
\end{figure}

This relation can be clearly demonstrated with the band diagram from Fig. \ref{Electrodes_band}(a),(b).
When the Fermi level in metal is inside the band gap of ferroelectric (Fig. \ref{Electrodes_band}(a)),
the energy Eq. \eqref{electrodes1} is positive, but it can be much smaller than in the case of isolated sample.
In this case the charged domain wall is metastable, but with the energy smaller than in the case of isolated sample.
If the fermi level in metal is above the bottom of the conduction band in ferroelectric (Fig. \ref{Electrodes_band}(b)), the energy Eq. \eqref{electrodes1} is negative.
The negative domain wall energy means that the configuration with "head-to-head" wall is stable, while in the case of  the isolated particle it is metastable.
The same result with some additional small terms for the sample with electrodes was obtained by Ivanchik, Chenskii et al.\cite{Guro70,Chenskii}

In the case of "tail-to-tail" wall, to provide screening, the electrons should move from the ferroelectric into the metal, the analog of the Eq. \eqref{electrodes1} can be written as follows:
\begin{equation}\label{electrodes2}
W=\frac{2P_0}{q}(E_\mathrm{F}-E_\mathrm{V}+A_\mathrm{f}-A_\mathrm{e})
\end{equation}

When the Fermi level in metal is inside the band gap of ferroelectric (Fig. \ref{Electrodes_band}(c)), the energy Eq. \eqref{electrodes2} is positive, and the charged domain wall is metastable.
If the fermi level in metal is below the top of the valence band in ferroelectric (Fig. \ref{Electrodes_band}(d)), the energy Eq. \eqref{electrodes2} is negative,
and the "tail-to-tail" domain wall is stable.

The preferable domain-wall configuration ("head-to-head" or "tail-to-tail") is defined by the direction of the electrons transfer, which depends on the work-function difference between the electrode and the ferroelectric.
This result is quite different from the case of an isolated sample, where "head-to-head" and "tail-to-tail" configurations have the same energy.

To take into account the surface energy term we should change $E_\mathrm{g}$ to $E_\mathrm{g}^\mathrm{eff}$ in Eqs. \eqref{electrodes1}, \eqref{electrodes2}, and the result will depend not only on the electron affinity difference, but also on the surface energy coefficient.
Depending on the sign of the electron affinity difference and the surface energy coefficient this effects can reinforce or work against each other.

\section{Numerical estimates and comparison with experimental data}\label{Experiment}

Let us compare the results obtained with experimental data.
First lets find what regime of screening corresponds to the typical perovskite materials at room temperature.
For BaTiO$_3$\cite{Wang} with $P_0=0.26$ C/m$^2\approx7.8\times10^4$ cgse and $|\alpha|=3.6\times10^{-3}$ the ferroelectric energy density $|\alpha|P_0^2=2.2\times10^7$ erg/cm$^3$.
 For PbTiO$_3$\cite{Fesenko73} with $P_0=0.75$ C/m$^2\approx2.2\times10^5$ cgse and $|\alpha|=0.05$ one can calculate $|\alpha|P_0^2=2.6\times10^9$ erg/cm$^3$.
 Both of this values are more than the critical value from Eq. \eqref{lim:degNl} for room temperature $\left( 2kT\right)^{5/2}m^{3/2}/(3\pi^2\hbar^3)=1.8\times10^6$ erg/cm$^3$.
 This means that the gas inside the domain wall is  degenerate.
 Typically the homogeneous carrier concentrations are non-degenerate.
Thus, for typical perovskites, screening is provided by the degenerate gas in the nonlinear regime.
The domain wall half-width can be found from Eq. \eqref{delta:degnl} and $\delta_\mathrm{deg}^\mathrm{nl}=16$ nm for BaTiO$_3$ and $\delta_\mathrm{deg}^\mathrm{nl}=2.7$ nm for PbTiO$_3$.
The corresponding typical concentration in the domain wall $n_\mathrm{c}\approx 10^{20}$ cm$^{-3}$ for BaTiO$_3$ and $n_\mathrm{c}\approx 1.7\times10^{21}$ cm$^{-3}$ for PbTiO$_3$.
These concentrations is still $2-3$ orders  smaller than the half-band-fill concentration and consistent with assumption we did to use the parabolic approximation.
At such concentrations the domain walls should have metallic conductivity\cite{Vul}.
Interesting, no experimental data of enhanced conductivity in "head-to-head" domain wall is available.
In contrast, in the recent experiment by Seidel et al. \cite{Ramesh} the neutral domain wall show enhanced conductivity, while the charged domain wall in the same sample does not.

\begin{figure}[h]
\begin{center}
\includegraphics[width=0.9\textwidth]{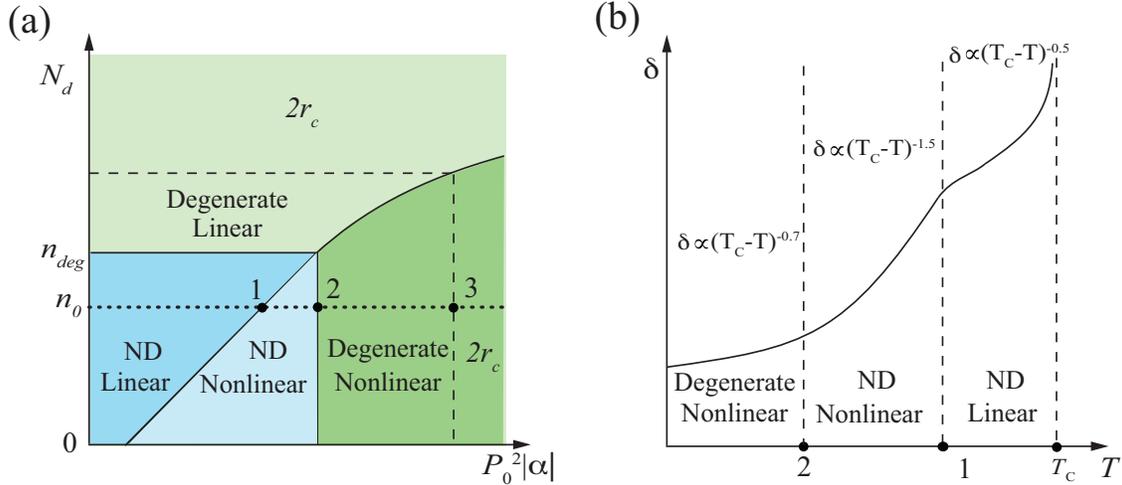}
\caption{(a) A diagram presenting areas that correspond to the different regimes of screening in a charged domain-wall in a large ferroelectric crystal, depending on the ferroelectric energy density $|\alpha|P_0^2$  and the concentration of the fully ionized dopant. The dotted line corresponds to the fixed homogeneous concentration of carriers. Approaching the ferroelectric phase transition we can move from right to left along the dotted line. (b) Schematic of temperature dependence of domain wall width for temperatures below $T_c$.}
\label{diagram_small}
\end{center}
\end{figure}

We can also check, what conditions correspond to different regimes of screening.
We use parameters of lead titanate \cite{Fesenko85} with $n_0=10^{18}$ cm$^{-3}$ and temperature about 700 K (close to PbTiO$_3$ phase transition) to calculate the spontaneous polarization, corresponding to different regimes of screening.
The dotted line on Fig. \ref{diagram_small}(a) represents our material at different spontaneous polarization, corresponding to different temperatures.
If the polarization is small enough, i.e. the corresponding point is on the left from the point 1 in Fig. \ref{diagram_small}(a), the regime of screening will be linear with classical electron gas.
From Eq. \eqref{lim:nondeg} we can calculate that the spontaneous polarization corresponding to point 1 is about 8 $\mu$C/cm$^2$.
Such values are possible close to the phase transition not in a bulk but in a thin film of PbTiO$_3$, where transition expected to be of the second order \cite{Pertsev}.
From Eq. \eqref{lim:degNl} and Eq. \eqref{lim:Prc} one can find that point 2 in Fig.\ref{diagram_small}(a) corresponds to $P_0=30$ $\mu$C/cm$^2$ and point 3 to $P_0=2400$ $\mu$C/cm$^2$.
The last two values are independent of $n_0$, we only assume that the homogenous concentration is not degenerate.
Point 3 can not be reached at any realistic spontaneous polarization, thus for typical materials charged domain wall is always wider than the neutral domain wall.
This fact confirms assumption we did in Sect. \ref{Analytical_Solutions}.
The "head-to-head" domain wall with the thickness about the correlation radius may be possible in a materials with metallic carrier concentration, like GeTe \cite{Moriguchi}.

Approaching phase transition we can pass through different regimes of screening, and this will be accompanying with the domain wall widening.
The dependences of domain wall width from the temperature difference $T_c-T$ are different for different regimes of screening and can be presented as $\delta\propto (T_c-T)^{-n}$, where $n=0.7$ for nonlinear screening with degenerate gas, $n=1.5$ and $n=0.5$ for nonlinear and linear screening with classical gas, respectively.
Schematically this dependence is shown in Fig. \ref{diagram_small}(b).

We can evaluate the ratio between domain wall half-widths of charged and neutral domain walls using Eq. ~\eqref{delta:degnl}:
\begin{equation}\label{ratio}
\frac{\delta_\mathrm{deg}^\mathrm{nl}}{\delta_\mathrm{neutral}}\approx\frac{3(a_0^3q)^{1/5}}{P_0^{1/5}\kappa^{1/2}|\alpha|^{1/10}}.
\end{equation}
where $a_0$ is the Bohr radius.
For BaTiO$_3$ this leads to the ratio $\delta_\mathrm{deg}^\mathrm{nl}/\delta_\mathrm{neutral}\approx 10$.
Equation \eqref{ratio} is only slightly dependent on the parameters of the material,
so that this ratio is virtually universal:
charged domain wall width for typical perovskite ferroelectrics not too close to the phase transition should be about one order larger than the width of the neutral domain wall.
This result is in good agreement with the experimental result of Jia et al.\cite{Jia}
where "head-to-head" and neutral domain walls widths were measured in a PZT thin films,
using the negative spherical-aberration imaging technique.
The width of a "head-to-head" domain wall is found to be about 10 unit cells, and width of neutral domain wall is about 1 unit cell.
As was already discussed, the results for the width of domain wall obtained in \ref{Analytical_Solutions}
is valid not only for isolated samples, but also for a sample with electrodes, thus we can compare it with the measurements done on thin films.

We can also compare the energy of uncharged wall and the energy of charged wall, neglecting polarizing effect of the surface, in an isolated crystal at room temperature.
The typical values of domain wall  energies for different types of domain walls in BaTiO$_3$ and PbTiO$_3$ are presented in the Tab. \ref{table:energies}.
The data for uncharged walls for BaTiO$_3$ is taken from Zhirnov\cite{Zhirnov}, Hlinka and Marton\cite{Hlinka}, and Bulaevskii\cite{Bulaevskii}.
For neutral domain wall of PbTiO$_3$ we used results of first-principles calculation by Poykko and Chadi\cite{Poykko} and Meyer and Vanderbilt,\cite{Meyer} (at $0$ K the obtained neutral domain wall energy is about $100$ erg/cm$^2$)
extrapolated to the room temperature using dependence $W_\mathrm{neutral}\propto(T-T_\mathrm{c})^{3/2}$ (see e.g. \cite{Tagantsev_book}).
To calculate the values for charged walls in BaTiO$_3$ we use $P_0=0.26$ C/m$^2\approx7.8\times10^4$ cgse and $E_\mathrm{g}=3.2$ eV.
For PbTiO$_3$ the spontaneous polarization $P_0=0.75$ C/m$^2\approx2.2\times10^5$ cgse and $E_\mathrm{g}=3.4$ eV.
\begin{table}
\caption{Rough estimates for the  domain wall formation energies BaTiO$_3$ and  PbTiO$_3$ at room temperature. The estimates for the charge walls do not take into account the surface poling effect.}  

\centering
\begin{tabular}{c c c c}
\hline\hline %inserts double horizontal lines
Material & Neutral & Charged  \\ [0.5ex]

\hline
BaTiO$_3$ & 10 & 1600 \\
PbTiO$_3$ & 25 & 5000 \\[1ex]

\hline %inserts single line
\end{tabular}
\label{table:energies}
\end{table}

Now we can estimate minimal thickness at which the charged domain wall is energetically favorable in comparison with the paraelectric state.
From Eq. \eqref{fe_der_para_vs_domains2}, without poling effect of the surface, one can find for room temperature for BaTiO$_3$ $L_\mathrm{par}=300$ $\mathrm{nm}$ and for PbTiO$_3$ $L_\mathrm{par}=80$ $\mathrm{nm}$.
For the samples with thickness $L<L_\mathrm{par}$, zero polarization is favorable.
Thus the existence of charged walls parallel to the film surfaces in thin films with the thickness about $10$ nm (like e.g. in \cite{Jia}) is possible only in the presence of injecting electrodes or strong polarizing effect of the surfaces.

If the poling effect of the surface is small, the charged domain wall can be formed as a result of the domain growth kinetic at phase transition.
The competing state, that can also occur as a result of such growing is the lamella structure with neutral domain walls (see Sect. \ref{Size}).
To estimate minimal thickness, that multidomain lamella structure is energetically favorable in comparison with the charged domain wall at phase transition we should find energy of neutral domain wall near phase transition.
It was shown that for second order phase transition $W_\mathrm{neutral}\propto P_0^3$.\cite{Mitsui}
We will use the same dependence to set a rough estimate for the neutral domain wall energy for the case of first order phase transition .\cite{Striffer}
The spontaneous polarization at phase transition is about $0.16$ C/m$^2$ for BaTiO$_3$ \cite{Wemple} and  $0.4$ C/m$^2$ for PbTiO$_3$.\cite{Fesenko85}
Using domain wall energies at room temperature (see Tab.~\ref{table:energies}) we estimate the neutral domain wall energies at phase transition $W_\mathrm{neutral}\approx2.5$ erg/cm$^2$ and $W_\mathrm{neutral}\approx4$ erg/cm$^2$ respectively.
Using Eq. \eqref{condition_multi}, neglecting poling ability of the surface, one can find for BaTiO$_3$ at the phase transition with $E_\mathrm{g}=3.2$ eV,
$\varepsilon_a\approx 5000$, $\varepsilon_c\approx 2500$\cite{Meyerhofer}  $L_\mathrm{md}=500$ $\mu$m.
For PbTiO$_3$ at the phase transition with $E_\mathrm{g}=3.4$ eV, $\varepsilon_a\approx 1500$,
$\varepsilon_c\approx 450$\cite{Haun} Eq. \eqref{condition_multi}  leads to much smaler values $L_\mathrm{md}=80$ $\mu$m.
This means that for thickness from $80$ $\mu$m to $500$ $\mu$m the charged domain wall is favorable in lead titanate, while in barium titanate laminar domain structure with neutral domain walls is.
This estimate corroborates with the experimental results obtained by Surowiak et al.\cite{Surowiak}
Indeed, the structures formed during the phase transition in lead titanate and barium titanate crystals with thickness about $100$ $\mathrm{ \mu m}$ were studied. In the case of lead titanate "head-to-head" domain walls were observed, whereas in barium titanate an a-c domains laminar structure was observed where c-domains are split into 180-degrees domains with neutral walls.
One should also mention that the lower conductivity of barium titanate can be also a factor preventing the formation of "head-to-head" domain walls.
Although in the case of nonlinear screening the homogeneous charge concentration does not affect the energy of the "head-to-head" configuration,
the screening free carriers should be delivered to the domain wall and the higher conductivity can assist in the formation of the charged domain wall.

In the case of the sample with electrodes, interesting observation could be done for example for BaTiO$_3$ with platinum electrodes. \cite{Vul,Ivanchik}
Neglecting poling effect of the surface for the case of the BaTiO$_3$ with platinum electrodes the formation energy of "tail-to tail" domain wall is negative, i.e. domain wall is energetically favorable in this case.
Indeed, neglecting poling effect of the surface Eq. \eqref{electrodes2} can be rewritten in a terms of band gap $E_\mathrm{g}$ and electron affinity $\chi$ of ferroelectric  in the form $W=(2P_0/q)(E_\mathrm{g}+\chi-A_\mathrm{e})$.
Using $E_\mathrm{g}=3.2$ eV, work function for platinum $A_\mathrm{e}=5.2$ eV, $\chi=0.4-1.2$ eV\cite{Morgulis,Busch} one can find $E_\mathrm{g}+\chi-A_\mathrm{e}=-(0.4 - 1.6)$ eV.
This situation have not been observed experimentally.
The possible explanation for this fact is the polarizing surface effect which works against the electron work-function difference.
In the considered case of "tail-to-tail" wall this means the preferable direction of polarization due to the surface effect is out of the surface.

\section{Conclusions}\label{conclusions}

The full set of equations suitable for numerical calculations of a "head-to-head" domain wall in an isolated sample was considered.
The effective boundary conditions for the case of the normal ferroelectrics (with $\varepsilon_\mathrm{b}/\varepsilon_\mathrm{f}\ll1$) was introduced.
Depending on spontaneous polarization and carriers concentration, different  regimes of screening and different spatial scales of charged domain walls are singled out.
For typical perovskites, not very close to the phase transition, this scale is about the nonlinear Thomas-Fermi screening length.
For perovskite ferroelectrics this scale is about one order larger than the width of a neutral domain wall, which is in the good agreement with experimental results for PZT.
Approaching phase transition by heating the sample we can pass through different regimes of screening; this will be accompanying with an appreciable widening of charged domain wall .
The dependence of domain wall width from the difference between the temperature and the curie temperature are different for different regimes of screening.

\begin{table}[h]
\caption{Summary of charged domain walls characteristics for different regimes of screening in isolated sample (without polarizing surface effect)}
\centering
\begin{tabularx}{\textwidth}{p{2.5cm} p{4cm} p{3.5cm} p{6cm}}
\hline\hline %inserts double horizontal lines
Regime & \centering Typical scale of domain wall half-width & \centering Formation energy per unit area of domain wall & \centering Remarks  \tabularnewline [0.5ex]

\hline
Linear \newline  classical &  \centering $\delta_\mathrm{cl}=\sqrt{\frac{2kT}{q^2(n_\mathrm{e0}+n_\mathrm{h0})|\alpha|}}$ & \centering $\frac{4}{3}|\alpha|P_0^2\delta_\mathrm{cl}$ &
Half-width is about the Debye screening length for a linear media with $\varepsilon_\mathrm{f}\approx2\pi/|\alpha|$ \\[1ex]

Nonlinear \newline classical &  \centering $\delta_\mathrm{cl}^\mathrm{nl}=\frac{4kT}{q|\alpha|P_0}$ & \centering $\frac{2P_0}{q}E_\mathrm{g}$ &
Half-width is about the nonlinear Debye screening length for a linear media with $\varepsilon_\mathrm{f}\approx2\pi/|\alpha|$\\[1ex]

Linear \newline degenerate &  \centering $\delta_\mathrm{deg}=\sqrt{\frac{2(3\pi ^2)^{2/3}\hbar ^2}{3mq^2n_0^{1/3}|\alpha|}}$ & \centering $\frac{4}{3}|\alpha|P_0^2\delta_\mathrm{deg}$ &
Half-width is about the Thomas-Fermi screening length for a linear media with $\varepsilon_\mathrm{f}\approx2\pi/|\alpha|$\\[1ex]

Nonlinear \newline degenerate &  \centering $\delta_\mathrm{deg}^\mathrm{nl}=\left(\frac{9\pi^4\hbar ^6}{q^5m^3|\alpha|^3P_0}\right)^{1/5}$ & \centering $\frac{2P_0}{q}E_\mathrm{g}$&
Suitable for the typical perovskites at room temperature. Half-width is about the nonlinear Thomas-Fermi screening length for a linear media with $\varepsilon_\mathrm{f}\approx2\pi/|\alpha|$\\[1ex]

\hline\hline %inserts single line
\end{tabularx}
\label{table:summary}
\end{table}

Expressions for a domain wall formation energy at different regimes of screening was obtained.
The information on the polarization profiles and domain wall energies without surface energy consideration for different possible screening regimes is summarized in Tab. \ref{table:summary}.
In the nonlinear regimes of screening this energy is equal to the energy necessary to create enough electron-hole pairs to fully screen the spontaneous polarization, being proportional to the electronic band gap of the ferroelectric.
The results obtained are closely related to the problem of internal screening for monodomain state.
It was shown that the energy of this screening is much higher than the energy obtained earlier by Ivanchik\cite{Ivanchik} and Watanabe\cite{Watanabe} due to neglect by the later an important contribution to it.
In general, the surface contribution to the formation energy of charged wall can be comparable with the formation energy obtained neglecting this effect.
Either "head-to-head" or "tail-to-tail" domain wall can be favorable in comparison with the internally screened monodomain state if surface energy is big enough.
If the charged domain wall is unfavorable, metastable "head-to-head" and "tail-to-tail" walls can be a result of the kinetic of domain growing.
As a result of such growth either "head-to-head" domains or multidomain state can be obtained.
Passing a second order phase transition, the multidomain state is always favorable, but for a first order phase transition, the favorable state depends on the thickness of the sample.

\section{Acknowledgments}\label{Acknowledgments}

This project was supported by the Swiss National Science Foundation. A.K.T. acknowledges Prof. Arkadii Levanyuk for a discussion of some issues addressed in this parer.

%APPENDIX
\appendix

\section{Boundary conditions.}\label{appendix}
To check the self-consistency of the scale $\delta$, we should check that we can neglect first term in Eq. \eqref{poissin_E}, i.e.
\begin{equation}\label{Self_cons}
\left|\varepsilon_\mathrm{b}\frac{\partial E}{\partial x}\right| \ll \left|4\pi \frac{\partial P}{\partial x}\right|.
\end{equation}
using Eq. \eqref{state} this inequality can be rewritten as:
\begin{equation}\label{Self_cons1}
\left|\left(\alpha+3\beta P^2\right) \frac{\partial P}{\partial x}-\kappa\frac{\partial^3 P}{\partial x^3}\right| \ll \left|\frac{4\pi}{\varepsilon_\mathrm{b}} \frac{\partial P}{\partial x}\right|.
\end{equation}
In solutions we obtained $P\leq P_0$, thus $3\beta P^2\leq 3|\alpha|P^2$ and using $\varepsilon_\mathrm{f}\gg\varepsilon_\mathrm{b}$ we can get:
\begin{equation}\label{Self_cons2}
\left|\left(\alpha+3\beta P^2\right) \frac{\partial P}{\partial x}\right| \leq \left|4\alpha \frac{\partial P}{\partial x}\right|
= \left|\frac{8\pi}{\varepsilon_\mathrm{f}} \frac{\partial P}{\partial x}\right|\ll \left|\frac{4\pi}{\varepsilon_\mathrm{b}} \frac{\partial P}{\partial x}\right|.
\end{equation}
To satisfy Eq. \eqref{Self_cons} we should also show
\begin{equation}\label{Self_cons3}
\left|\kappa\frac{\partial^3 P}{\partial x^3}\right| \ll \left|\frac{4\pi}{\varepsilon_\mathrm{b}} \frac{\partial P}{\partial x}\right|.
\end{equation}
For the case of linear screening, using solutions $P_0=\tanh(x/\delta)$, one can get:
\begin{equation}\label{ratio_linear}
\frac{\kappa\varepsilon_\mathrm{b}}{4\pi}\frac{\partial^3 P/\partial x^3}{ \partial P/\partial x}=\frac{\kappa\varepsilon_\mathrm{b}}{2\pi\delta^2}\left[\frac{2\sinh^2(x/\delta)-1}{1+\sinh^2(x/\delta)}\right].
\end{equation}
The maximum value of the function in parenthesis is $2$, thus, using $\delta\geq 2r_\mathrm{c}$:
\begin{equation}\label{ratio_linear1}
\frac{\kappa\varepsilon_\mathrm{b}}{4\pi}\frac{\partial^3 P/\partial x^3}{ \partial P/\partial x}=
\frac{\varepsilon_\mathrm{b}}{\varepsilon_\mathrm{f}}\left(\frac{2r_\mathrm{c}}{\delta}\right)^2\leq\frac{\varepsilon_\mathrm{b}}{\varepsilon_\mathrm{f}}\ll1.
\end{equation}
In the case of nonlinear screening with the classical gas from solution \eqref{ClasNL:exsol}:
\begin{equation}\label{ratio_clas_nl}
\frac{\kappa\varepsilon_\mathrm{b}}{4\pi}\frac{\partial^3 P/\partial x^3}{ \partial P/\partial x}=
\frac{4\kappa\varepsilon_\mathrm{b}}{\pi\delta^2}\left| (1-p^2)^2(1-7p^2) \right|,
\end{equation}
where $-1<p<1$. The maximum value of the function in modulus is 1, thus:
\begin{equation}\label{ratio_clas_nl1}
\frac{\kappa\varepsilon_\mathrm{b}}{4\pi}\frac{\partial^3 P/\partial x^3}{ \partial P/\partial x}\leq
\frac{4\kappa\varepsilon_\mathrm{b}}{\pi\delta^2}=\frac{4\varepsilon_\mathrm{b}}{\varepsilon_\mathrm{f}}\left(\frac{2r_\mathrm{c}}{\delta}\right)^2\leq\frac{4\varepsilon_\mathrm{b}}{\varepsilon_\mathrm{f}}\ll1.
\end{equation}
In the case of nonlinear screening with degenerate electron gas using solution \eqref{degen_nonlinear_exact}, one can find:
\begin{equation}\label{ratio_clas_nl1}
\frac{\kappa\varepsilon_\mathrm{b}}{4\pi}\frac{\partial^3 P/\partial x^3}{ \partial P/\partial x}<
\frac{\kappa\varepsilon_\mathrm{b}}{4\pi\delta^2}=\frac{\varepsilon_\mathrm{b}}{\varepsilon_\mathrm{f}}\left(\frac{r_\mathrm{c}}{\delta}\right)^2<\frac{\varepsilon_\mathrm{b}}{\varepsilon_\mathrm{f}}\ll1.
\end{equation}
\section{Free energy density for screening by classical and degenerate electron gas.}\label{appendix1}

In this section we will calculate the kinetic energy contribution in the free energy density $\Phi_\mathrm{eg}(n)-\Phi_\mathrm{eg}(n_0)$.

In the case of the classical electron gas, from Eq. \eqref{entropy} for $f_\mathrm{C}(p)\ll1$, the main term in the entropy density in the conduction band can be written in the following form:
\begin{equation}\label{classentropy_conduction1}
S_\mathrm{eg}^\mathrm{C}=-k\sum\limits_i \left[f_i^\mathrm{C}\ln f_i^\mathrm{C}- f_i^\mathrm{C}\right],
\end{equation}
where summation is done over all possible states in the conduction band.
The Fermi function for a classical gas has the following form:
\begin{equation}\label{class_conduction}
f_i^\mathrm{C}=\exp\left(-\frac{E_i-E_\mathrm{F}-q\varphi}{kT}\right),
\end{equation}
where $E_i$ is the energy of the state.
Together with Eq. \eqref{classentropy_conduction1} this leads to
\begin{equation}\label{classentropy_conductive}
S_\mathrm{eg}^\mathrm{C}=k\sum\limits_i\left[\left(\frac{E_i-E_\mathrm{F}-q\varphi}{kT}+1\right)f_i^\mathrm{C}\right]
\end{equation}

The energy density in the conduction band can be presented as follows:
\begin{equation}\label{classenergy_conductive}
E_\mathrm{eg}^\mathrm{C}=\sum\limits_iE_if_i^\mathrm{C}.
\end{equation}

Thus, the kinetic free energy density of classical electron gas in the conduction band can be found as follows:
\begin{equation}\label{classfreeenergy_conductive}
\Phi_\mathrm{eg}^\mathrm{C}=E_\mathrm{eg}^\mathrm{C}-TS_\mathrm{eg}^\mathrm{C}=\sum\limits_i\left( E_\mathrm{F}+q\varphi-kT\right)f_i^\mathrm{C}=\left( E_\mathrm{F}+q\varphi-kT\right)n_\mathrm{e}
\end{equation}

On the same lines, the free energy density for classical electron gas in the valence band can be found in the form:
\begin{equation}\label{classfreeenergy_valence}
\Phi_\mathrm{eg}^\mathrm{V}=-\left( E_\mathrm{F}+q\varphi+kT\right)n_\mathrm{h}
\end{equation}
and the total kinetic free energy is equal to the sum of \eqref{classfreeenergy_conductive} and \eqref{classfreeenergy_valence} and can be presented in the following form:
\begin{equation}\label{classfreeenergy}
\Phi_ \mathrm{eg}=\left( E_\mathrm{F}+q\varphi\right)(n_\mathrm{e}-n_\mathrm{h})-kT(n_\mathrm{e}+n_\mathrm{h})
\end{equation}

Note, that Eq. \eqref{classfreeenergy} presents only the kinetic part of free energy, the electrostatic energy is already taken into account separately as $\varepsilon_\mathrm{b}E^2/8\pi$ in Eq. \eqref{fe}.
The Fermi level is constant,
thus in a sample with fixed number of electrons, Eq. \eqref{classfreeenergy}  can be rewritten as follows:
\begin{equation}\label{classfreeenergy_1}
\Phi_ \mathrm{eg}=q\varphi(n_\mathrm{e}-n_\mathrm{h})-kT(n_\mathrm{e}+n_\mathrm{h})
\end{equation}

For degenerate Fermi gas, as was mentioned in section \ref{Degenerate}, two situations are possible: either the valence band is fully occupied and there are electrons in the conduction band, or there are holes in the valence band and the conduction band is empty.
With parabolic approximation the energy density can be calculated directly from Eq. ~\eqref{energy_bandgap} in terms of concentration, in a form which is universal for both situations as follows:
\begin{equation}\label{gasenergyDeg}
E_\mathrm{eg} = \frac{3\hbar ^2(3\pi ^2)^{2/3}}{10m}(n_\mathrm{e}^{5/3}+n_\mathrm{h}^{5/3})+n_\mathrm{e}E_\mathrm{g}.
\end{equation}

The main term in the entropy density of a degenerate gas with respect to $T/T_\mathrm{F}$, where $T_\mathrm{F}$ is the Fermi temperature,
can be found from Eq.~\eqref{entropy} as follows \cite{Landau5} :
\begin{equation}\label{entropyDeg}
S_\mathrm{eg} = \left(\frac{\pi}{3}\right)^{2/3}\frac{m}{\hbar ^2}k^2Tn^{1/3},
\end{equation}
where $n$ signifies either the concentration of electrons or the concentration of holes.

The condition of applicability of the strongly degenerate Fermi gas approximation requires the temperature to be small in comparison to the Fermi temperature:
\begin{equation}\label{degen}
kT \ll \frac{\hbar ^2}{m}n^{2/3}.
\end{equation}
Thus, using Eq. \eqref{gasenergyDeg}
\begin{equation}\label{TS}
TS_\mathrm{eg}\ll \frac{\hbar ^4}{m^2}n^{4/3}\left(\frac{\pi}{3}\right)^{2/3}\frac{m}{\hbar ^2}n^{1/3}=\frac{3\hbar ^2(3\pi ^2)^{2/3}}{10m}n^{5/3}\leq E_\mathrm{eg},
\end{equation}
and in Eq.~\eqref{freeenergy} we can neglect the term with entropy in comparison with the energy term.
Thus, the free energy of the degenerate electron gas can be found from \eqref{gasenergyDeg} as follows:
\begin{equation}\label{feDeggas}
\Phi_\mathrm{eg} =E_\mathrm{eg} =  \frac{3\hbar ^2(3\pi ^2)^{2/3}}{10m}(n_\mathrm{e}^{5/3}+n_\mathrm{h}^{5/3})+n_\mathrm{e}E_\mathrm{g}.
\end{equation}

So that for the degenerate gas we can finally get:
\begin{equation}\label{feDeggas1}
\Phi_\mathrm{eg}(n)-\Phi_\mathrm{eg}(n_0)=\frac{3\hbar ^2(3\pi ^2)^{2/3}}{10m}(n_\mathrm{e}^{5/3}+n_\mathrm{h}^{5/3}-n_\mathrm{e0}^{5/3}-n_\mathrm{h0}^{5/3})+(n_\mathrm{e}-n_\mathrm{e0})E_\mathrm{g}.
\end{equation}

%APPENDIX 2
\section{Calculation of the energy of a sample with "head-to-head" domain walls.}\label{appendix2}

%%%%%%%%%%%%%%%%%%%%%%%%%%%%%%%%%%%%%%%%%%%%%
\subsection{Linear screening by classical electron gas.}\label{appendix2_CL_lin}

For a classical gas, free energy density is given by Eq. \eqref{energy_classicaleg}. Using \eqref{concentration:class}, the electron gas part of the free energy can be presented as a function of $\partial P/\partial x$:
\begin{equation}\label{classfreeenergy_gas_1}
\Phi_\mathrm{eg}(x)-\Phi_\mathrm{eg0}=q\varphi\left(n_\mathrm{e0}-n_\mathrm{h0}-\frac{1}{q}\frac{\partial P}{\partial x}\right)-kT\left(n_\mathrm{e0}e^{\frac{q\varphi}{kT}}+n_\mathrm{h0}e^{-\frac{q\varphi}{kT}}\right)
\end{equation}
where the potential $\varphi$ can be found from  Eq. \eqref{classic_exponent}
\begin{equation}\label{potential_linear}
\varphi=\frac{kT}{q}\ln\left[\frac{1}{2n_\mathrm{e0}}\left(-\frac{1}{q}\frac{\partial P}{\partial x}+n_\mathrm{e0}-n_\mathrm{h0}+\sqrt{\left(-\frac{1}{q}\frac{\partial P}{\partial x}+n_\mathrm{e0}-n_\mathrm{h0}\right)^2+4n_\mathrm{e0}n_\mathrm{h0}}\right)\right].
\end{equation}
In the case of linear screening, Eq. \eqref{classfreeenergy_gas_1} can be presented as an expansion with respect to a small parameter $\theta=\frac{1}{q(n_\mathrm{e0}+n_\mathrm{h0})}\frac{\partial P}{\partial x}$ and the second order terms should be taken into account:
\begin{equation}\label{classfreeenergy_gas_2}
\Phi_\mathrm{eg}(x)-\Phi_\mathrm{eg0}=A\times\theta+\frac{1}{2}kT(n_\mathrm{e0}+n_\mathrm{h0})\theta^2+o(\theta^2)
\end{equation}
where $A$ is some coefficient independent of $\theta$, $o(\theta^2)$ designates the higher order terms with respect to $\theta^2$.

Now we will obtain the relation between the ferroelectric and the electron gas parts of the free energy.
With the change of variables:
\begin{equation}\label{varchange_t}
t=\frac{\partial P}{\partial x}
\end{equation}
Eq. \eqref{classiclargen} can be rewritten in the following form:
\begin{equation}\label{classic_t}
\alpha P+\beta P^3=\frac{kT}{q^2(n_\mathrm{e0}+n_\mathrm{h0})}t\frac{\partial t}{\partial P}.
\end{equation}
Integrating this equation once in $P$ and taking into account that in the center of the domain $t=0$ and $P=P_0$, we can get:
\begin{equation}\label{energy_ratio_CL}
\frac{\alpha P^2}{2}+\frac{\beta P^4}{4}-\frac{\alpha P_0^2}{4}=\frac{1}{2}\frac{kT}{q^2(n_\mathrm{e0}+n_\mathrm{h0})}\left(\frac{\partial P}{\partial x}\right)^2.
\end{equation}
Using Eqs. \eqref{energy_classicaleg},  \eqref{energy_ratio_CL} and second order term from Eq. \eqref{classfreeenergy_gas_2} (first order terms will give zero after integration along the thickness of the sample with the boundary conditions, Eq. \eqref{Boundary_P}), one can obtain:
\begin{equation}\label{classic_lin_energy_hh}
\int_{-\infty}^{\infty}[ \Phi_\mathrm{h-h}(x)-\Phi_0(x)]dx=2\int_{-\infty}^{\infty}\left( \frac{\alpha P^2}{2}+\frac{\beta P^4}{4}-\frac{\alpha P_0^2}{4}\right)dx.
\end{equation}
From Eq. \eqref{energy_ratio_CL}, $\frac{\partial P}{\partial x}$ can be presented as follows:
\begin{equation}\label{dPdx_CL_Lin}
\frac{\partial P}{\partial x}=\sqrt{\left(\frac{\alpha P^2}{2}+\frac{\beta P^4}{4}-\frac{\alpha P_0^2}{4}\right)\frac{2q^2(n_\mathrm{e0}+n_\mathrm{h0})}{kT}}.
\end{equation}
With Eq. \eqref{dPdx_CL_Lin}, the integral in Eq. \eqref{classic_lin_energy_hh} can be rewritten as an integral over the polarization as follows:
\begin{equation}\label{classic_lin_energy_hh1}
\int_{-\infty}^{\infty}[ \Phi_\mathrm{h-h}(x)-\Phi_0(x)]dx=2\int_{-P_0}^{P_0}\sqrt{\left( \frac{\alpha P^2}{2}+\frac{\beta P^4}{4}-\frac{\alpha P_0^2}{4}\right)\frac{kT}{2q^2(n_\mathrm{e0}+n_\mathrm{h0})}}dP=\frac{2}{3}|\alpha|P_0^2\delta_\mathrm{cl}
\end{equation}
In the case of linear screening by classical electron gas the polarization distribution for "head-to-head" and "tail-to-tail" walls are the same to within the sign, thus both terms in Eq. \eqref{W_separate} are equal, and the total energy of a charged wall can be given as follows:
\begin{equation}\label{W_cl_lin}
W_\mathrm{cl}=\frac{4}{3}|\alpha|P_0^2\delta_\mathrm{cl}
\end{equation}

%%%%%%%%%%%%%%%%%%%%%%%%%%%%%%%%%%%%%%%%%%%%%%%%%%%%%%

\subsection{Nonlinear screening by classical electron gas.}\label{appendix2_CL_NL}

In the case of nonlinear screening, using Eq. \eqref{Poisson_P}, Eq. \eqref{classfreeenergy_1} can be presented in the following form:
\begin{equation}\label{classfreeenergy_2}
\Phi(x)-\Phi_0=\frac{\alpha}{2}P^2+\frac{\beta}{4}P^4-\frac{\alpha}{4}P_0^2-\varphi\left(\frac{\partial P}{\partial x}\right)-\frac{kT}{q}\left|\frac{\partial P}{\partial x}\right|
\end{equation}

In the case of nonlinear screening by classical electron gas, Eq. \eqref{classicNL} (with the variable change given by Eq. \eqref{varchange_t}) can be rewritten in the following form:
\begin{equation}\label{classicNL_t}
\alpha P+\beta P^3=-\frac{kT}{q}\frac{\partial t}{\partial P}
\end{equation}
Integrating this equation once, as we did for the case of the linear screening, we will find the ferroelectric part of the free energy density in the following form:
\begin{equation}\label{energy_ratio_CL_NL}
\frac{\alpha P^2}{2}+\frac{\beta P^4}{4}-\frac{\alpha P_0^2}{4}=-\frac{kT}{q}\left(\frac{\partial P}{\partial x}\right)
\end{equation}
Then, the ferroelectric part of the energy of the sample, associated with the domain wall region can be found as follows:
\begin{equation}\label{classic_NL_FE_energy}
W_\mathrm{fe}=\int_{-\infty}^{\infty}-\frac{kT}{q}\left(\frac{\partial P}{\partial x}\right)dx=\int_{P_0}^{-P_0}-\frac{kT}{q}dP=\frac{2P_0kT}{q}=\frac{1}{2}|\alpha|P_0^2\delta_\mathrm{cl}^\mathrm{nl}
\end{equation}
From Eq. \eqref{concentration:class}, in the case of the nonlinear screening, the potential can be presented in the following form:
\begin{equation}\label{potential_NL}
\varphi=\frac{kT}{q}\ln \left(-\frac{\partial P/\partial x}{qn_\mathrm{e0}}\right)
\end{equation}
and the kinetic free energy of electron gas in Eq. \eqref{energy_classicaleg} can be rewritten in the form:
\begin{equation}\label{classic_NL_kin_energy}
W_\mathrm{eg}=-\frac{kT}{q}\int_{-\infty}^{\infty}\left(\ln\left(- \frac{\partial P/\partial x}{qn_\mathrm{e0}}\right)-1\right)\left(\frac{\partial P}{\partial x}\right)dx=\frac{kT}{q}\int_{-P_0}^{P_0}\left(\ln\left(- \frac{\partial P/\partial x}{qn_\mathrm{e0}}\right)-1\right)dP
\end{equation}
With $\partial P/\partial x$ coming from Eq. \eqref{energy_ratio_CL_NL}, Eq. \eqref{classic_NL_kin_energy} yields:
\begin{equation}\label{classic_NL_kin_energy1}
W_\mathrm{eg}=\frac{2kTP_0}{q}\left(\ln\left(\frac{4|\alpha|P_0^2}{kTn_\mathrm{e0}}\right)-5\right)
\end{equation}
Thus, the first term in \eqref{W_separate} can be obtained as a sum of Eqs. \eqref{classic_NL_FE_energy} and \eqref{classic_NL_kin_energy1} as follows:
\begin{equation}\label{classic_NL_h_h}
W_\mathrm{h-h}=\frac{2kTP_0}{q}\left(\ln\left(\frac{4|\alpha|P_0^2}{kTn_\mathrm{e0}}\right)-4\right)
\end{equation}

On the same lines, for the  "tail-to-tail" domain wall region, one can obtain the following energy:
\begin{equation}\label{classic_NL_t_t}
W_\mathrm{t-t}=\frac{2kTP_0}{q}\left(\ln\left(\frac{4|\alpha|P_0^2}{kTn_\mathrm{h0}}\right)-4\right)
\end{equation}

The energy of the sample with charged wall can be obtained as a sum of Eqs. \eqref{classic_NL_h_h} and \eqref{classic_NL_t_t}, taking into account the mass action law $n_\mathrm{i}^2=n_\mathrm{e0}n_\mathrm{h0}$:
\begin{equation}\label{W_cl_nl1}
W_\mathrm{cl}^\mathrm{nl}=\frac{4kTP_0}{q}\left(\ln\left(\frac{4|\alpha|P_0^2}{kTn_\mathrm{i}}\right)-4\right)
\end{equation}
Using $n_\mathrm{i}=Ne^{-E_\mathrm{g}/2kT}$, where $N=\sqrt{N_\mathrm{C}N_\mathrm{V}}$, Eq. \eqref {W_cl_nl1} can be rewritten in the following form:
\begin{equation}\label{W_cl_nl2}
W_\mathrm{cl}^\mathrm{nl}=\frac{2P_0}{q}E_\mathrm{g}+\frac{4kTP_0}{q}\left(\ln\left(\frac{4|\alpha|P_0^2}{kTN}\right)-4\right)
\end{equation}
The second term in Eq. \eqref{W_cl_nl2} can be neglected in comparison with the first one, and the domain wall energy can be written as:
\begin{equation}\label{W_cl_nl3}
W_\mathrm{cl}^\mathrm{nl}=\frac{2P_0}{q}E_\mathrm{g}
\end{equation}
Indeed, Eq. \eqref{W_cl_nl2} can be rewritten in the form:
\begin{equation}\label{W_cl_nl4}
W_\mathrm{cl}^\mathrm{nl}=\frac{2P_0}{q}\left(E_\mathrm{g}+2kT\ln\left(\frac{4|\alpha|P_0^2}{kTN}\right)-8kT\right)
\end{equation}
Typically in semiconductors $E_\mathrm{g}$ is about several electron-volts and $kT$ at room temperature is about $0.026$ eV, thus $8kT$ can be neglected in comparison with $E_\mathrm{g}$.
 Now we can estimate the term with logarithm in Eq. \eqref{W_cl_nl2} in comparison with $E_\mathrm{g}$.
The density of states $N$ can be estimated as follows\cite{Sze}:
\begin{equation}\label{N_estimation}
N\approx 2\left(\frac{mkT}{2\pi\hbar^2}\right)^{3/2}
\end{equation}
The conditions for nonlinear screening by classical gas, corresponding to line 7 and to lines 1 and 2 in Fig.\ref{diagram2} can be written as:
\begin{equation}\label{condition_NL_cl_7}
|\alpha|P_0^2\ll\frac{(kT)^{5/2}m^{3/2}}{\hbar^3}
\end{equation}
and
\begin{equation}\label{condition_NL_cl_12}
n_\mathrm{i}\leq n_0\ll\frac{|\alpha|P_0^2}{kT},
\end{equation}
respectively.
Using Eqs. \eqref{condition_NL_cl_7}, \eqref{condition_NL_cl_12},  we can find following estimates:
\begin{equation}\label{est_log_max}
2kT\ln\left(\frac{4|\alpha|P_0^2}{kTN}\right)\ll2kT\ln(2\pi^2)\approx 6kT\ll E_\mathrm{g}.
\end{equation}
\begin{equation}\label{est_log_min}
2kT\ln\left(\frac{4|\alpha|P_0^2}{kTN}\right)\gg-2kT\ln\left(\frac{4n_\mathrm{i}}{N}\right)>-E_\mathrm{g}.
\end{equation}
Equations \eqref{est_log_max} and \eqref{est_log_min} imply:
\begin{equation}\label{est_log_min}
\left|2kT\ln\left(\frac{4|\alpha|P_0^2}{kTN}\right)\right|\ll E_\mathrm{g}.
\end{equation}
%

%%%%%%%%%%%%%%%%%%%%%%%%%%%%%%%%%%%%%%%%%%%%%
\subsection{Linear screening by degenerate electron gas.}\label{appendix2_deg_lin}

In the case of a linear screening by degenerate electron gas, the amount of the electrons in the conduction bands remains constant and the carriers are just redistributed inside the band (either the valence band is fully occupied, and the electrons are redistributed into the conduction band, or the conduction band is empty and the holes are redistributed into the valence band). Thus the electron gas energy can be found from Eq. \eqref{feDeggas1} as follows:
\begin{equation}\label{deg_lin_kin_ener1}
\Phi_\mathrm{eg}-\Phi_\mathrm{eg0}=\frac{3\hbar ^2(3\pi ^2)^{2/3}}{10m}\left( \left(n_0+\frac{1}{q} \frac{\partial P}{\partial x}\right)^{5/3}-n_0^{5/3}\right)
\end{equation}
This equation can be expanded with respect to the small parameter $\frac{1}{qn_0} \frac{\partial P}{\partial x}$ and the linear term can be neglected, because it will give zero after integration through the sample thickness due to the boundary conditions \eqref{Boundary_P}.
Thus, Eq. \eqref{deg_lin_kin_ener1} can be presented in the form
\begin{equation}\label{deg_lin_kin_ener2}
\Phi_\mathrm{eg}-\Phi_\mathrm{eg0}=\frac{\hbar ^2(3\pi ^2)^{2/3}}{6mn_0^{1/3}}\left(  \frac{\partial P}{\partial x}\right)^2
\end{equation}
Integrating Eq. \eqref{degen_lin}, one can calculate the local relation between the ferroelectric part of the energy and the part associated with the electron gas in the following form:
\begin{equation}\label{deg_lin_relation}
\frac{\alpha P^2}{2}+\frac{\beta P^4}{4}-\frac{\alpha P_0^2}{4}=\frac{\hbar ^2(3\pi ^2)^{2/3}}{6mn_0^{1/3}}\left(  \frac{\partial P}{\partial x}\right)^2
\end{equation}
In the linear regime of screening, the polarization distributions for the "head-to-head" and "tail-to-tail" domain walls are the same to within the sign, and with Eqs.  \eqref{feDeggas1}, \eqref{deg_lin_kin_ener2} and \eqref{deg_lin_relation} domain wall energy can be found in the following form:
\begin{equation}\label{W_deg_lin1}
W_\mathrm{deg}=4\int_{-\infty}^{\infty}\left(\frac{\alpha P^2}{2}+\frac{\beta P^4}{4}-\frac{\alpha P_0^2}{4}\right)dx
\end{equation}

With Eq. \eqref{deg_lin_relation}, the integral can be rewritten as an integral by polarization, and the domain-wall energy can be found as follows:
\begin{equation}\label{W_deg_lin2}
W_\mathrm{deg}=4\int_{-P_0}^{P_0}\sqrt{\frac{\alpha P^2}{2}+\frac{\beta P^4}{4}-\frac{\alpha P_0^2}{4}}\sqrt{\frac{\hbar ^2(3\pi ^2)^{2/3}}{6mn_0^{1/3}}}dP=\frac{4}{3}|\alpha|P_0^2\delta_\mathrm{deg}
\end{equation}

%%%%%%%%%%%%%%%%%%%%%%%%%%%%%%%%%%%%%%%%%%%%%
\subsection{Nonlinear screening by degenerate electron gas.}\label{appendix2_deg_NL}

In the most important case of nonlinear screening by degenerate gas, integrating Eq. \eqref{degen_nonlinear} the following relation can be obtained:
\begin{equation}\label{deg_nl_relation}
\frac{\alpha P^2}{2}+\frac{\beta P^4}{4}-\frac{\alpha P_0^2}{4}=\frac{\hbar ^2(3\pi ^2)^{2/3}}{5mq^{5/3}}\left(  \frac{\partial P}{\partial x}\right)^{5/3}
\end{equation}

From Eqs. \eqref{feDeggas} and \eqref{deg_nl_relation}, taking into account that $\int n_\mathrm{e} dx=2P_0/q$, the energy of the domain wall can be found as follows:
\begin{equation}\label{W_deg_nl_a1}
W_\mathrm{deg}^\mathrm{nl}=5\int_{-\infty}^{\infty}\left(\frac{\alpha P^2}{2}+\frac{\beta P^4}{4}-\frac{\alpha P_0^2}{4}\right)dx+\frac{2P_0}{q}E_\mathrm{g}
\end{equation}
where the polarization profile is obtained for the domain wall with the boundary conditions of Eq. \eqref{boundary1}.
As we did for the other type of walls, the integral in Eq. \eqref{W_deg_nl_a1} using Eq. \eqref{deg_nl_relation} can be presented as an integral by $P$ as:
\begin{equation}\label{W_deg_nl_a2}
\int_{-P_0}^{P_0}\left(\frac{\alpha P^2}{2}+\frac{\beta P^4}{4}-\frac{\alpha P_0^2}{4}\right)\left(\frac{\hbar ^2(3\pi ^2)^{2/3}}{5mq^{5/3}}\right)^{3/5}dP=\frac{\int_0^1(1-z^2)^{4/5}dz}{4^{2/5}5^{3/5}}|\alpha|P_0^2\delta_\mathrm{deg}^\mathrm{nl}
\end{equation}
and Eq. \eqref{W_deg_nl_a1} can be rewritten in a following form:
\begin{equation}\label{W_deg_nl_a3}
W_\mathrm{deg}^\mathrm{nl}=\left(\frac{5}{4}\right)^{2/5}\int_0^1(1-z^2)^{4/5}dz\times|\alpha|P_0^2\delta_\mathrm{deg}^\mathrm{nl}+\frac{2P_0}{q}E_\mathrm{g}\approx0.77\times |\alpha|P_0^2\delta_\mathrm{deg}^\mathrm{nl}+\frac{2P_0}{q}E_\mathrm{g}.
\end{equation}
The first term in Eq. \eqref{W_deg_nl_a3} can be neglected if
\begin{equation}\label{W_deg_nl_a4}
P_0\ll=\frac{1}{\sqrt{|\alpha|}}E_\mathrm{g}^{5/4}\left(\frac{m}{\hbar^2}\right)^{3/4}<E_\mathrm{g}^{5/4}\left(\frac{m}{\hbar^2}\right)^{3/4}\approx10^6\,\mathrm{cgse}\approx300\, \mathrm{\mu C/cm^2}.
\end{equation}
Where we supposed $E_\mathrm{g}\approx 3$ $ \mathrm{eV}$. Thus the energy of the domain wall can be written as follows:

\begin{equation}\label{W_deg_nl_a3}
W_\mathrm{deg}^\mathrm{nl}=\frac{2P_0}{q}E_\mathrm{g}
\end{equation}

%\begin{thebibliography}{999}
%\bibliography{Isolated2}

\begin{thebibliography}{10}%
\makeatletter
\providecommand \@ifxundefined [1]{%
 \ifx #1\undefined \expandafter \@firstoftwo
 \else \expandafter \@secondoftwo
\fi
}%
\providecommand \@ifnum [1]{%
 \ifnum #1\expandafter \@firstoftwo
 \else \expandafter \@secondoftwo
\fi
}%
\providecommand \enquote [1]{``#1''}%
\providecommand \bibnamefont  [1]{#1}%
\providecommand \bibfnamefont [1]{#1}%
\providecommand \citenamefont [1]{#1}%
\providecommand\href[0]{\@sanitize\@href}%
\providecommand\@href[1]{\endgroup\@@startlink{#1}\endgroup\@@href}%
\providecommand\@@href[1]{#1\@@endlink}%
\providecommand \@sanitize [0]{\begingroup\catcode`\&12\catcode`\#12\relax}%
\@ifxundefined \pdfoutput {\@firstoftwo}{%
 \@ifnum{\z@=\pdfoutput}{\@firstoftwo}{\@secondoftwo}%
}{%
 \providecommand\@@startlink[1]{\leavevmode}%
 \providecommand\@@endlink[0]{}%
}{%
 \providecommand\@@startlink[1]{%
  \leavevmode
  \pdfstartlink
   attr{/Border[0 0 1 ]/H/I/C[0 1 1]}%
   user{/Subtype/Link/A<</Type/Action/S/URI/URI(#1)>>}%
  \relax
 }%
 \providecommand\@@endlink[0]{\pdfendlink}%
}%
\providecommand \url  [0]{\begingroup\@sanitize \@url }%
\providecommand \@url [1]{\endgroup\@href {#1}{\urlprefix}}%
\providecommand \urlprefix [0]{URL }%
\providecommand \Eprint[0]{\href }%
\@ifxundefined \urlstyle {%
  \providecommand \doi [1]{doi:\discretionary{}{}{}#1}%
}{%
  \providecommand \doi [0]{doi:\discretionary{}{}{}\begingroup
  \urlstyle{rm}\Url }%
}%
\providecommand \doibase [0]{http://dx.doi.org/}%
\providecommand \Doi[1]{\href{\doibase#1}}%
\providecommand \bibAnnote [3]{%
  \BibitemShut{#1}%
  \begin{quotation}\noindent
    \textsc{Key:}\ #2\\\textsc{Annotation:}\ #3%
  \end{quotation}%
}%
\providecommand \bibAnnoteFile [2]{%
  \IfFileExists{#2}{\bibAnnote {#1} {#2} {\input{#2}}}{}%
}%
\providecommand \typeout [0]{\immediate \write \m@ne }%
\providecommand \selectlanguage [0]{\@gobble}%
\providecommand \bibinfo [0]{\@secondoftwo}%
\providecommand \bibfield [0]{\@secondoftwo}%
\providecommand \translation [1]{[#1]}%
\providecommand \BibitemOpen[0]{}%
\providecommand \bibitemStop [0]{}%
\providecommand \bibitemNoStop [0]{.\EOS\space}%
\providecommand \EOS [0]{\spacefactor3000\relax}%
\providecommand \BibitemShut [1]{\csname bibitem#1\endcsname}%
%</preamble>
\bibitem{Setter}%
  \BibitemOpen
  \bibfield{author}{%
  \bibinfo {author} {\bibfnamefont{N.}~\bibnamefont{Setter}}\ and\ \bibinfo
  {author} {\bibfnamefont{R.}~\bibnamefont{Waser}},\ }%
  \bibfield{journal}{%
  \bibinfo {journal} {Acta materialia}\ }%
  \textbf{\bibinfo {volume} {48}},\ \bibinfo {pages} {151} (\bibinfo {year}
  {2000})%
  \bibAnnoteFile{NoStop}{Setter}%
\bibitem{Scott}%
  \BibitemOpen
  \bibfield{author}{%
  \bibinfo {author} {\bibfnamefont{J.~F.}\ \bibnamefont{Scott}},\ }%
  \bibfield{journal}{%
  \bibinfo {journal} {Science}\ }%
  \textbf{\bibinfo {volume} {315}},\ \bibinfo {pages} {954} (\bibinfo {year}
  {2007})%
  \bibAnnoteFile{NoStop}{Scott}%
\bibitem{Levanyuk}%
  \BibitemOpen
  \bibfield{author}{%
  \bibinfo {author} {\bibfnamefont{B.~A.}\ \bibnamefont{Strukov}}\ and\
  \bibinfo {author} {\bibfnamefont{A.~P.}\ \bibnamefont{Levanyuk}},\ }%
  \emph{\bibinfo {title} {Ferroelectric Phenomena in Crystals}}\ (\bibinfo
  {publisher} {Springer},\ \bibinfo {year} {1998})\ Chap.\ \bibinfo {chapter}
  {3.2, 10.2}%
  \bibAnnoteFile{NoStop}{Levanyuk}%
\bibitem{Fesenko73}%
  \BibitemOpen
  \bibfield{author}{%
  \bibinfo {author} {\bibfnamefont{E.~G.}\ \bibnamefont{Fesenko}}, \bibinfo
  {author} {\bibfnamefont{V.~G.}\ \bibnamefont{Gavrilyatchenko}}, \bibinfo
  {author} {\bibfnamefont{M.~A.}\ \bibnamefont{Martinenko}}, \bibinfo {author}
  {\bibfnamefont{A.~F.}\ \bibnamefont{Semenchov}},\ and\ \bibinfo {author}
  {\bibfnamefont{I.~P.}\ \bibnamefont{Lapin}},\ }%
  \bibfield{journal}{%
  \bibinfo {journal} {Ferroelectrics}\ }%
  \textbf{\bibinfo {volume} {6}},\ \bibinfo {pages} {61} (\bibinfo {year}
  {1973})%
  \bibAnnoteFile{NoStop}{Fesenko73}%
\bibitem{Surowiak}%
  \BibitemOpen
  \bibfield{author}{%
  \bibinfo {author} {\bibfnamefont{Z.}~\bibnamefont{Surowiak}}, \bibinfo
  {author} {\bibfnamefont{J.}~\bibnamefont{Dec}}, \bibinfo {author}
  {\bibfnamefont{R.}~\bibnamefont{Skulski}}, \bibinfo {author}
  {\bibfnamefont{E.~G.}\ \bibnamefont{Fesenko}}, \bibinfo {author}
  {\bibfnamefont{V.~G.}\ \bibnamefont{Gavrilyatchenko}},\ and\ \bibinfo
  {author} {\bibfnamefont{A.~F.}\ \bibnamefont{Semenchov}},\ }%
  \bibfield{journal}{%
  \bibinfo {journal} {Ferroelectrics}\ }%
  \textbf{\bibinfo {volume} {20}},\ \bibinfo {pages} {277} (\bibinfo {year}
  {1978})%
  \bibAnnoteFile{NoStop}{Surowiak}%
\bibitem{Fesenko85}%
  \BibitemOpen
  \bibfield{author}{%
  \bibinfo {author} {\bibfnamefont{E.~G.}\ \bibnamefont{Fesenko}}, \bibinfo
  {author} {\bibfnamefont{V.~G.}\ \bibnamefont{Gavrilyatchenko}}, \bibinfo
  {author} {\bibfnamefont{A.~F.}\ \bibnamefont{Semenchov}},\ and\ \bibinfo
  {author} {\bibfnamefont{S.~M.}\ \bibnamefont{Yufatova}},\ }%
  \bibfield{journal}{%
  \bibinfo {journal} {Ferroelectrics}\ }%
  \textbf{\bibinfo {volume} {63}},\ \bibinfo {pages} {289} (\bibinfo {year}
  {1985})%
  \bibAnnoteFile{NoStop}{Fesenko85}%
\bibitem{Randall}%
  \BibitemOpen
  \bibfield{author}{%
  \bibinfo {author} {\bibfnamefont{C.~A.}\ \bibnamefont{Randall}}\ and\
  \bibinfo {author} {\bibfnamefont{D.}~\bibnamefont{Barber}},\ }%
  \bibfield{journal}{%
  \bibinfo {journal} {Journal of Materials Science}\ }%
  \textbf{\bibinfo {volume} {22}},\ \bibinfo {pages} {925} (\bibinfo {year}
  {1987})%
  \bibAnnoteFile{NoStop}{Randall}%
\bibitem{Ramesh}%
Analysis of Fig. 1 in
  \BibitemOpen
  \bibfield{author}{%
  \bibinfo {author} {\bibfnamefont{J.}~\bibnamefont{Seidel}}, \bibinfo {author}
  {\bibfnamefont{L.~W.}\ \bibnamefont{Martin}}, \bibinfo {author}
  {\bibfnamefont{Q.}~\bibnamefont{He}}, \bibinfo {author}
  {\bibfnamefont{Q.}~\bibnamefont{Zhan}}, \bibinfo {author}
  {\bibfnamefont{Y.-H.}\ \bibnamefont{Chu}}, \bibinfo {author}
  {\bibfnamefont{A.}~\bibnamefont{Rother}}, \bibinfo {author}
  {\bibfnamefont{P.}~\bibnamefont{Hawkridge},
  \bibfnamefont{M.~E.~Maksymovych}}, \bibinfo {author}
  {\bibfnamefont{P.}~\bibnamefont{Yu}}, \bibinfo {author}
  {\bibfnamefont{M.}~\bibnamefont{Gajek}}, \bibinfo {author}
  {\bibfnamefont{N.}~\bibnamefont{Balke}}, \bibinfo {author}
  {\bibfnamefont{S.~V.}\ \bibnamefont{Kalinin}}, \bibinfo {author}
  {\bibfnamefont{S.}~\bibnamefont{Gemming}}, \bibinfo {author}
  {\bibfnamefont{F.}~\bibnamefont{Wang}}, \bibinfo {author}
  {\bibfnamefont{G.}~\bibnamefont{Catalan}}, \bibinfo {author}
  {\bibfnamefont{J.~F.}\ \bibnamefont{Scott}}, \bibinfo {author}
  {\bibfnamefont{N.}~\bibnamefont{Spaldin}}, \bibinfo {author}
  {\bibfnamefont{J.}~\bibnamefont{Orenstein}},\ and\ \bibinfo {author}
  {\bibfnamefont{R.}~\bibnamefont{Ramesh}},\ }%
  \bibfield{journal}{%
  \bibinfo {journal} {Nature Materials}\ }%
  \textbf{\bibinfo {volume} {8}},\ \bibinfo {pages} {229} (\bibinfo {year}
  {2009})%
   \ show that 71$^\circ$ domain wall is a "tail-to-tail" domain wall
  \bibAnnoteFile{NoStop}{Ramesh}%
\bibitem{Jia}%
  \BibitemOpen
  \bibfield{author}{%
  \bibinfo {author} {\bibfnamefont{C.-L.}\ \bibnamefont{Jia}}, \bibinfo
  {author} {\bibfnamefont{S.-B.}\ \bibnamefont{Mi}}, \bibinfo {author}
  {\bibfnamefont{K.}~\bibnamefont{Urban}}, \bibinfo {author}
  {\bibfnamefont{I.}~\bibnamefont{Vrejoiu}}, \bibinfo {author}
  {\bibfnamefont{M.}~\bibnamefont{Alexe}},\ and\ \bibinfo {author}
  {\bibfnamefont{D.}~\bibnamefont{Hesse}},\ }%
  \bibfield{journal}{%
  \bibinfo {journal} {Nature Materials}\ }%
  \textbf{\bibinfo {volume} {7}},\ \bibinfo {pages} {57} (\bibinfo {year}
  {2008})%
  \bibAnnoteFile{NoStop}{Jia}%
\bibitem{Mokry}%
  \BibitemOpen
  \bibfield{author}{%
  \bibinfo {author} {\bibfnamefont{P.}~\bibnamefont{Mokr\'{y}}}, \bibinfo
  {author} {\bibfnamefont{A.~K.}\ \bibnamefont{Tagantsev}},\ and\ \bibinfo
  {author} {\bibfnamefont{J.}~\bibnamefont{Fousek}},\ }%
  \bibfield{journal}{%
  \bibinfo {journal} {Phys. Rev. B}\ }%
  \textbf{\bibinfo {volume} {75}},\ \bibinfo {pages} {094110} (\bibinfo {year}
  {2007})%
  \bibAnnoteFile{NoStop}{Mokry}%
\bibitem{Kalinin}%
  \BibitemOpen
  \bibfield{author}{%
  \bibinfo {author} {\bibfnamefont{N.}~\bibnamefont{Balke}}, \bibinfo {author}
  {\bibfnamefont{M.}~\bibnamefont{Gajek}}, \bibinfo {author}
  {\bibfnamefont{A.~K.}\ \bibnamefont{Tagantsev}}, \bibinfo {author}
  {\bibfnamefont{L.~W.}\ \bibnamefont{Martin}}, \bibinfo {author}
  {\bibfnamefont{Y.-H.}\ \bibnamefont{Chu}}, \bibinfo {author}
  {\bibfnamefont{R.}~\bibnamefont{Ramesh}},\ and\ \bibinfo {author}
  {\bibfnamefont{S.~V.}\ \bibnamefont{Kalinin}},\ }%
  \bibfield{journal}{%
  \bibinfo {journal} {Advanced Functional Materials}\ }%
  \textbf{\bibinfo {volume} {20}},\ \bibinfo {pages} {3466} (\bibinfo {year}
  {2010})%
  \bibAnnoteFile{NoStop}{Kalinin}%
\bibitem{Chenskii}%
  \BibitemOpen
  \bibfield{author}{%
  \bibinfo {author} {\bibfnamefont{V.~F.}\ \bibnamefont{Krapivin}}\ and\
  \bibinfo {author} {\bibfnamefont{E.~V.}\ \bibnamefont{Chenskii}},\ }%
  \bibfield{journal}{%
  \bibinfo {journal} {Sov. Phys. - Solid State}\ }%
  \textbf{\bibinfo {volume} {12}},\ \bibinfo {pages} {454} (\bibinfo {year}
  {1970})%
  \bibAnnoteFile{NoStop}{Chenskii}%
\bibitem{Guro68}%
  \BibitemOpen
  \bibfield{author}{%
  \bibinfo {author} {\bibfnamefont{G.~M.}\ \bibnamefont{Guro}}, \bibinfo
  {author} {\bibfnamefont{I.~I.}\ \bibnamefont{Ivanchik}},\ and\ \bibinfo
  {author} {\bibfnamefont{N.~F.}\ \bibnamefont{Kovtonyuk}},\ }%
  \bibfield{journal}{%
  \bibinfo {journal} {Sov. Phys. - Solid State}\ }%
  \textbf{\bibinfo {volume} {10}},\ \bibinfo {pages} {100} (\bibinfo {year}
  {1968})%
  \bibAnnoteFile{NoStop}{Guro68}%
\bibitem{Guro70}%
  \BibitemOpen
  \bibfield{author}{%
  \bibinfo {author} {\bibfnamefont{G.~M.}\ \bibnamefont{Guro}}, \bibinfo
  {author} {\bibfnamefont{I.~I.}\ \bibnamefont{Ivanchik}},\ and\ \bibinfo
  {author} {\bibfnamefont{N.~F.}\ \bibnamefont{Kovtonyuk}},\ }%
  \bibfield{journal}{%
  \bibinfo {journal} {Sov.Phys. - Solid State}\ }%
  \textbf{\bibinfo {volume} {11}},\ \bibinfo {pages} {1574} (\bibinfo {year}
  {1970})%
  \bibAnnoteFile{NoStop}{Guro70}%
\bibitem{Vul}%
  \BibitemOpen
  \bibfield{author}{%
  \bibinfo {author} {\bibfnamefont{B.~M.}\ \bibnamefont{Vul}}, \bibinfo
  {author} {\bibfnamefont{G.~M.}\ \bibnamefont{Guro}},\ and\ \bibinfo {author}
  {\bibfnamefont{I.~I.}\ \bibnamefont{Ivanchik}},\ }%
  \bibfield{journal}{%
  \bibinfo {journal} {Ferroelectrics}\ }%
  \textbf{\bibinfo {volume} {6}},\ \bibinfo {pages} {29} (\bibinfo {year}
  {1973})%
  \bibAnnoteFile{NoStop}{Vul}%
\bibitem{Yudin}%
  \BibitemOpen
  \bibfield{author}{%
  \bibinfo {author} {\bibfnamefont{S.~P.}\ \bibnamefont{Yudin}}, \bibinfo
  {author} {\bibfnamefont{T.~V.}\ \bibnamefont{Panchenko}},\ and\ \bibinfo
  {author} {\bibfnamefont{K.~A.}\ \bibnamefont{Yu.}},\ }%
  \bibfield{journal}{%
  \bibinfo {journal} {Ferroelectrics}\ }%
  \textbf{\bibinfo {volume} {18}},\ \bibinfo {pages} {45} (\bibinfo {year}
  {1978})%
  \bibAnnoteFile{NoStop}{Yudin}%
\bibitem{Vanderbilt}%
  \BibitemOpen
  \bibfield{author}{%
  \bibinfo {author} {\bibfnamefont{X.}~\bibnamefont{Wu}}\ and\ \bibinfo
  {author} {\bibfnamefont{D.}~\bibnamefont{Vanderbilt}},\ }%
  \bibfield{journal}{%
  \bibinfo {journal} {Phys. Rev. B}\ }%
  \textbf{\bibinfo {volume} {73}},\ \bibinfo {pages} {020103} (\bibinfo {year}
  {2006})%
  \bibAnnoteFile{NoStop}{Vanderbilt}%
\bibitem{Fridkin}%
  \BibitemOpen
  \bibfield{author}{%
  \bibinfo {author} {\bibfnamefont{V.~M.}\ \bibnamefont{Fridkin}},\ }%
  \emph{\bibinfo {title} {Ferroelectric semiconductors}}\ (\bibinfo {publisher}
  {Colsuntants bureau},\ \bibinfo {address} {New York},\ \bibinfo {year}
  {1980})\ Chap.\ \bibinfo {chapter} {3.2}%
  \bibAnnoteFile{NoStop}{Fridkin}%
\bibitem{Tagantsev_book}%
  \BibitemOpen
  \bibfield{author}{%
  \bibinfo {author} {\bibfnamefont{A.~A.}\ \bibnamefont{Tagantsev}}, \bibinfo
  {author} {\bibfnamefont{L.~E.}\ \bibnamefont{Cross}},\ and\ \bibinfo {author}
  {\bibfnamefont{J.}~\bibnamefont{Fousek}},\ }%
  \emph{\bibinfo {title} {Domains in Ferroic Crystals and Thin Films}}\
  (\bibinfo {publisher} {Springer},\ \bibinfo {address} {New York},\ \bibinfo
  {year} {2010})\ Chap.\ \bibinfo {chapter} {5.2, 6.1}%
  \bibAnnoteFile{NoStop}{Tagantsev_book}%
\bibitem{Sze}%
  \BibitemOpen
  \bibfield{author}{%
  \bibinfo {author} {\bibfnamefont{S.~M.}\ \bibnamefont{Sze}},\ }%
  \emph{\bibinfo {title} {Semiconductor Devices Physics and Technology}}\
  (\bibinfo {publisher} {John Wiley and Sons},\ \bibinfo {year} {1985})\ Chap.\
  \bibinfo {chapter} {1.6}, p.~\bibinfo {pages} {18}%
  \bibAnnoteFile{NoStop}{Sze}%
\bibitem{Liu}%
  \BibitemOpen
  \bibfield{author}{%
  \bibinfo {author} {\bibfnamefont{Y.~Y.}\ \bibnamefont{Liu}}\ and\ \bibinfo
  {author} {\bibfnamefont{J.~Y.}\ \bibnamefont{Li}},\ }%
  \bibfield{journal}{%
  \bibinfo {journal} {Appl. Phys. Lett.}\ }%
  \textbf{\bibinfo {volume} {97}},\ \bibinfo {pages} {042905} (\bibinfo {year}
  {2010})%
  \bibAnnoteFile{NoStop}{Liu}%
\bibitem{Xiao}%
  \BibitemOpen
  \bibfield{author}{%
  \bibinfo {author} {\bibfnamefont{Y.}~\bibnamefont{Xiao}}, \bibinfo {author}
  {\bibfnamefont{V.~B.}\ \bibnamefont{Shenoy}},\ and\ \bibinfo {author}
  {\bibfnamefont{K.}~\bibnamefont{Bhattacharya}},\ }%
  \bibfield{journal}{%
  \bibinfo {journal} {Appl. Phys. Lett.}\ }%
  \textbf{\bibinfo {volume} {95}},\ \bibinfo {pages} {247603} (\bibinfo {year}
  {2005})%
  \bibAnnoteFile{NoStop}{Xiao}%
\bibitem{Glinchuk}%
  \BibitemOpen
  \bibfield{author}{%
  \bibinfo {author} {\bibfnamefont{M.~D.}\ \bibnamefont{Glinchuk}}\ and\
  \bibinfo {author} {\bibfnamefont{A.~N.}\ \bibnamefont{Morozovska}},\ }%
  \bibfield{journal}{%
  \bibinfo {journal} {J. Phys.: Condens. Matter}\ }%
  \textbf{\bibinfo {volume} {16}},\ \bibinfo {pages} {3517} (\bibinfo {year}
  {2004})%
  \bibAnnoteFile{NoStop}{Glinchuk}%
\bibitem{Minyukov}%
  \BibitemOpen
  \bibfield{author}{%
  \bibinfo {author} {\bibfnamefont{A.~P.}\ \bibnamefont{Levanyuk}}\ and\
  \bibinfo {author} {\bibfnamefont{S.~A.}\ \bibnamefont{Minyukov}},\ }%
  \bibfield{journal}{%
  \bibinfo {journal} {Fiz. Tverd. Tela}\ }%
  \textbf{\bibinfo {volume} {25}},\ \bibinfo {pages} {2617} (\bibinfo {year}
  {1983})%
  \bibAnnoteFile{NoStop}{Minyukov}%
\bibitem{Tagantsev}%
  \BibitemOpen
  \bibfield{author}{%
  \bibinfo {author} {\bibfnamefont{A.~K.}\ \bibnamefont{Tagantsev}}, \bibinfo
  {author} {\bibfnamefont{G.}~\bibnamefont{Gerra}},\ and\ \bibinfo {author}
  {\bibfnamefont{N.}~\bibnamefont{Setter}},\ }%
  \bibfield{journal}{%
  \bibinfo {journal} {Phys. Rev. B}\ }%
  \textbf{\bibinfo {volume} {77}},\ \bibinfo {pages} {174111} (\bibinfo {year}
  {2008})%
  \bibAnnoteFile{NoStop}{Tagantsev}%
\bibitem{Landau5}%
  \BibitemOpen
  \bibfield{author}{%
  \bibinfo {author} {\bibfnamefont{L.~D.}\ \bibnamefont{Landau}}\ and\ \bibinfo
  {author} {\bibfnamefont{E.~M.}\ \bibnamefont{Lifshits}},\ }%
  \emph{\bibinfo {title} {Statistical Physics}}\ (\bibinfo {publisher}
  {Butterworth-Heinemann},\ \bibinfo {address} {Oxford},\ \bibinfo {year}
  {1982})\ Chap.\ \bibinfo {chapter} {56, 57}%
  \bibAnnoteFile{NoStop}{Landau5}%
\bibitem{Landau6}%
  \BibitemOpen
  \bibfield{author}{%
  \bibinfo {author} {\bibfnamefont{L.~D.}\ \bibnamefont{Landau}}\ and\ \bibinfo
  {author} {\bibfnamefont{E.~M.}\ \bibnamefont{Lifshits}},\ }%
  \emph{\bibinfo {title} {Fluid Mechanics}}\ (\bibinfo {publisher} {Pergamon
  Press},\ \bibinfo {address} {Oxford},\ \bibinfo {year} {1987})\
  Chap.~\bibinfo {chapter} {39}%
  \bibAnnoteFile{NoStop}{Landau6}%
\bibitem{Tagantsev87}%
  \BibitemOpen
  \bibfield{author}{%
  \bibinfo {author} {\bibfnamefont{A.~K.}\ \bibnamefont{Tagantsev}}, \bibinfo
  {author} {\bibfnamefont{I.~G.}\ \bibnamefont{Sinii}},\ and\ \bibinfo {author}
  {\bibfnamefont{S.~D.}\ \bibnamefont{Prokhorova}},\ }%
  \bibfield{journal}{%
  \bibinfo {journal} {Izv. Akad. Nauk SSSR, Ser. Fiz.}\ }%
  \textbf{\bibinfo {volume} {51}},\ \bibinfo {pages} {2082} (\bibinfo {year}
  {1987})%
  \bibAnnoteFile{NoStop}{Tagantsev87}%
\bibitem{Ivanchik}%
  \BibitemOpen
  \bibfield{author}{%
  \bibinfo {author} {\bibfnamefont{I.~I.}\ \bibnamefont{Ivanchik}},\ }%
  \bibfield{journal}{%
  \bibinfo {journal} {Ferroelectrics}\ }%
  \textbf{\bibinfo {volume} {145}},\ \bibinfo {pages} {149} (\bibinfo {year}
  {1993})%
  \bibAnnoteFile{NoStop}{Ivanchik}%
\bibitem{Watanabe}%
  \BibitemOpen
  \bibfield{author}{%
  \bibinfo {author} {\bibfnamefont{Y.}~\bibnamefont{Watanabe}},\ }%
  \bibfield{journal}{%
  \bibinfo {journal} {Phys. Rev. B}\ }%
  \textbf{\bibinfo {volume} {57}},\ \bibinfo {pages} {789} (\bibinfo {year}
  {1998})%
  \bibAnnoteFile{NoStop}{Watanabe}%
\bibitem{Wang}%
  \BibitemOpen
  \bibfield{author}{%
  \bibinfo {author} {\bibfnamefont{Y.~L.}\ \bibnamefont{Wang}}, \bibinfo
  {author} {\bibfnamefont{A.~K.}\ \bibnamefont{Tagantsev}}, \bibinfo {author}
  {\bibfnamefont{D.}~\bibnamefont{Damjanovich}}, \bibinfo {author}
  {\bibfnamefont{N.}~\bibnamefont{Setter}}, \bibinfo {author}
  {\bibfnamefont{V.~K.}\ \bibnamefont{Yarmarkin}}, \bibinfo {author}
  {\bibfnamefont{A.~I.}\ \bibnamefont{Sokolov}},\ and\ \bibinfo {author}
  {\bibfnamefont{I.~A.}\ \bibnamefont{Lukyanchuk}},\ }%
  \bibfield{journal}{%
  \bibinfo {journal} {J. Appl. Phys.}\ }%
  \textbf{\bibinfo {volume} {101}},\ \bibinfo {pages} {104115} (\bibinfo {year}
  {2007})%
  \bibAnnoteFile{NoStop}{Wang}%
\bibitem{Pertsev}%
  \BibitemOpen
  \bibfield{author}{%
  \bibinfo {author} {\bibfnamefont{N.~A.}\ \bibnamefont{Pertsev}}, \bibinfo
  {author} {\bibfnamefont{A.~G.}\ \bibnamefont{Zembilgotov}},\ and\ \bibinfo
  {author} {\bibfnamefont{A.~K.}\ \bibnamefont{Tagantsev}},\ }%
  \bibfield{journal}{%
  \bibinfo {journal} {Phys. Rev. Lett.}\ }%
  \textbf{\bibinfo {volume} {80}},\ \bibinfo {pages} {1988} (\bibinfo {year}
  {1998})%
  \bibAnnoteFile{NoStop}{Pertsev}%
\bibitem{Moriguchi}%
  \BibitemOpen
  \bibfield{author}{%
  \bibinfo {author} {\bibfnamefont{Y.}~\bibnamefont{Moriguchi}}\ and\ \bibinfo
  {author} {\bibfnamefont{Y.}~\bibnamefont{Koga}},\ }%
  \bibfield{journal}{%
  \bibinfo {journal} {J. Phys. Soc. Jpn.}\ }%
  \textbf{\bibinfo {volume} {12}},\ \bibinfo {pages} {100} (\bibinfo {year}
  {1957})%
  \bibAnnoteFile{NoStop}{Moriguchi}%
\bibitem{Zhirnov}%
  \BibitemOpen
  \bibfield{author}{%
  \bibinfo {author} {\bibfnamefont{V.~A.}\ \bibnamefont{Zhirnov}},\ }%
  \bibfield{journal}{%
  \bibinfo {journal} {Sov. Phys. - JETP}\ }%
  \textbf{\bibinfo {volume} {8}},\ \bibinfo {pages} {822} (\bibinfo {year}
  {1959})%
  \bibAnnoteFile{NoStop}{Zhirnov}%
\bibitem{Hlinka}%
  \BibitemOpen
  \bibfield{author}{%
  \bibinfo {author} {\bibfnamefont{J.}~\bibnamefont{Hlinka}}\ and\ \bibinfo
  {author} {\bibfnamefont{P.}~\bibnamefont{M\'arton}},\ }%
  \bibfield{journal}{%
  \bibinfo {journal} {Phys. Rev. B}\ }%
  \textbf{\bibinfo {volume} {74}},\ \bibinfo {pages} {104104} (\bibinfo {year}
  {2006})%
  \bibAnnoteFile{NoStop}{Hlinka}%
\bibitem{Bulaevskii}%
  \BibitemOpen
  \bibfield{author}{%
  \bibinfo {author} {\bibfnamefont{L.~N.}\ \bibnamefont{Bulaevskii}},\ }%
  \bibfield{journal}{%
  \bibinfo {journal} {Sov. Phys. - Solid State}\ }%
  \textbf{\bibinfo {volume} {5}},\ \bibinfo {pages} {2329} (\bibinfo {year}
  {1964})%
  \bibAnnoteFile{NoStop}{Bulaevskii}%
\bibitem{Poykko}%
  \BibitemOpen
  \bibfield{author}{%
  \bibinfo {author} {\bibfnamefont{S.}~\bibnamefont{Poykko}}\ and\ \bibinfo
  {author} {\bibfnamefont{D.}~\bibnamefont{Chadi}},\ }%
  \bibfield{journal}{%
  \bibinfo {journal} {Appl. Phys. Lett.}\ }%
  \textbf{\bibinfo {volume} {21}},\ \bibinfo {pages} {2830} (\bibinfo {year}
  {1999})%
  \bibAnnoteFile{NoStop}{Poykko}%
\bibitem{Meyer}%
  \BibitemOpen
  \bibfield{author}{%
  \bibinfo {author} {\bibfnamefont{B.}~\bibnamefont{Meyer}}\ and\ \bibinfo
  {author} {\bibfnamefont{D.}~\bibnamefont{Vanderbilt}},\ }%
  \bibfield{journal}{%
  \bibinfo {journal} {Phys. Rev. B}\ }%
  \textbf{\bibinfo {volume} {65}},\ \bibinfo {pages} {104111} (\bibinfo {year}
  {2002})%
  \bibAnnoteFile{NoStop}{Meyer}%
\bibitem{Mitsui}%
  \BibitemOpen
  \bibfield{author}{%
  \bibinfo {author} {\bibfnamefont{T.}~\bibnamefont{Mitsui}}\ and\ \bibinfo
  {author} {\bibfnamefont{J.}~\bibnamefont{Furuichi}},\ }%
  \bibfield{journal}{%
  \bibinfo {journal} {Phys. Rev.}\ }%
  \textbf{\bibinfo {volume} {90}},\ \bibinfo {pages} {193} (\bibinfo {year}
  {1953})%
  \bibAnnoteFile{NoStop}{Mitsui}%
\bibitem{Striffer}%
  \BibitemOpen
  \bibfield{author}{%
  \bibinfo {author} {\bibfnamefont{S.~K.}\ \bibnamefont{Streiffer}}, \bibinfo
  {author} {\bibfnamefont{J.~A.}\ \bibnamefont{Eastman}}, \bibinfo {author}
  {\bibfnamefont{D.~D.}\ \bibnamefont{Fong}}, \bibinfo {author}
  {\bibfnamefont{C.}~\bibnamefont{Thompson}}, \bibinfo {author}
  {\bibfnamefont{A.}~\bibnamefont{Munkholm}}, \bibinfo {author}
  {\bibfnamefont{M.}~\bibnamefont{Ramana~Murty}}, \bibinfo {author}
  {\bibfnamefont{O.}~\bibnamefont{Auciello}}, \bibinfo {author}
  {\bibfnamefont{G.~R.}\ \bibnamefont{Bai}},\ and\ \bibinfo {author}
  {\bibfnamefont{G.~B.}\ \bibnamefont{Stephenson}},\ }%
  \bibfield{journal}{%
  \bibinfo {journal} {J. Appl. Phys.}\ }%
  \textbf{\bibinfo {volume} {101}},\ \bibinfo {pages} {104115} (\bibinfo {year}
  {2007})%
  \bibAnnoteFile{NoStop}{Striffer}%
\bibitem{Wemple}%
  \BibitemOpen
  \bibfield{author}{%
  \bibinfo {author} {\bibfnamefont{S.~H.}\ \bibnamefont{Wemple}}, \bibinfo
  {author} {\bibfnamefont{M.}~\bibnamefont{Didomenico}~\bibnamefont{Jr.}},\ and\ \bibinfo
  {author} {\bibfnamefont{I.}~\bibnamefont{Camlibel}},\ }%
  \bibfield{journal}{%
  \bibinfo {journal} {J. Phys. Chem. Solids}\ }%
  \textbf{\bibinfo {volume} {29}},\ \bibinfo {pages} {1797} (\bibinfo {year}
  {1968})%
  \bibAnnoteFile{NoStop}{Wemple}%
\bibitem{Meyerhofer}%
  \BibitemOpen
  \bibfield{author}{%
  \bibinfo {author} {\bibfnamefont{D.}~\bibnamefont{Meyerhofer}},\ }%
  \bibfield{journal}{%
  \bibinfo {journal} {Phys. Rev.}\ }%
  \textbf{\bibinfo {volume} {112}},\ \bibinfo {pages} {413} (\bibinfo {year}
  {1958})%
  \bibAnnoteFile{NoStop}{Meyerhofer}%
\bibitem{Haun}%
  \BibitemOpen
  \bibfield{author}{%
  \bibinfo {author} {\bibfnamefont{M.~J.}\ \bibnamefont{Haun}}, \bibinfo
  {author} {\bibfnamefont{E.}~\bibnamefont{Furman}}, \bibinfo {author}
  {\bibfnamefont{S.~J.}\ \bibnamefont{Jang}}, \bibinfo {author}
  {\bibfnamefont{H.~A.}\ \bibnamefont{McKinstry}},\ and\ \bibinfo {author}
  {\bibfnamefont{L.~E.}\ \bibnamefont{Cross}},\ }%
  \bibfield{journal}{%
  \bibinfo {journal} {J. Appl. Phys.}\ }%
  \textbf{\bibinfo {volume} {62}},\ \bibinfo {pages} {3331} (\bibinfo {year}
  {1987})%
  \bibAnnoteFile{NoStop}{Haun}%
\bibitem{Morgulis}%
  \BibitemOpen
  \bibfield{author}{%
  \bibinfo {author} {\bibfnamefont{N.~L.}\ \bibnamefont{Morgulis}},\ }%
  \bibfield{journal}{%
  \bibinfo {journal} {Sov. Phys. - Tech. Phys.}\ }%
  \textbf{\bibinfo {volume} {2}},\ \bibinfo {pages} {391} (\bibinfo {year}
  {1957})%
  \bibAnnoteFile{NoStop}{Morgulis}%
\bibitem{Busch}%
  \BibitemOpen
  \bibfield{author}{%
  \bibinfo {author} {\bibfnamefont{G.}~\bibnamefont{Busch}}, \bibinfo {author}
  {\bibfnamefont{H.}~\bibnamefont{Flury}},\ and\ \bibinfo {author}
  {\bibfnamefont{W.}~\bibnamefont{Merz}},\ }%
  \bibfield{journal}{%
  \bibinfo {journal} {Helv. Phys. Acta}\ }%
  \textbf{\bibinfo {volume} {21}},\ \bibinfo {pages} {212} (\bibinfo {year}
  {1948})%
  \bibAnnoteFile{NoStop}{Busch}%
\end{thebibliography}
%\bibliographystyle{unsrt}
%\nocite{*}
%\end{thebibliography}

%%%%%%%%%%%%%%%%
%\begin{comment}

%

%\end{comment}

\end{document}